\documentclass[superscriptaddress,amsmath,twocolumn,amssymb,prd,preprintnumbers,showpacs]{revtex4-1}
\usepackage{graphicx}
\usepackage{amssymb}
\usepackage{amsmath}
\usepackage{amsfonts}
\usepackage[dvipsnames]{xcolor}
\usepackage{float}
\usepackage[colorlinks=true,pdfstartview=FitV,linkcolor=blue,citecolor=blue,urlcolor=blue,breaklinks=true]{hyperref}%
\usepackage{upgreek}
\usepackage{cancel}
\usepackage{amsopn}
\usepackage{float}
\usepackage[cmtip,arrow]{xy}
\usepackage{pb-diagram,pb-xy}
\usepackage{slashed}
\usepackage{bbold}
\usepackage{amsopn}
\usepackage{fancyhdr}
\usepackage[yyyymmdd,hhmmss]{datetime}
\providecommand{\U}[1]{\protect\rule{.1in}{.1in}}
\newcommand{\be}{\begin{equation}}
\newcommand{\ee}{\end{equation}}
\newcommand{\bea}{\begin{eqnarray}}
\newcommand{\eea}{\end{eqnarray}}
\newcommand{\ba}{\begin{eqnarray*}}
\newcommand{\ea}{\end{eqnarray*}}

\newcommand{\q}[1]{``#1''}

\def\normOrd#1{\mathop{:}\nolimits\!#1\!\mathop{:}\nolimits}

\begin{document}

\title{Causality, unitarity, and indefinite metric in Maxwell-Chern-Simons extensions}

%\date{\today}

\author{Ricardo Avila}
\email{raavilavi@gmail.com}
\affiliation{Departamento de Ciencias B\'{a}sicas, Universidad del B\'{\i}o-B\'{\i}o, 3800708, Chill\'{a}n, Chile}
\author{Jose R. Nascimento}
\email{jroberto@fisica.ufpb.br}
\affiliation{Departamento de F\'{\i}sica, Universidade Federal da Para\'{\i}ba, 58051-970, Jo\~ao Pessoa, Para\'{\i}ba, Brazil}
\author{Albert Yu. Petrov}
\email{petrov@fisica.ufpb.br}
\affiliation{Departamento de F\'{\i}sica, Universidade Federal da Para\'{\i}ba, 58051-970, Jo\~ao Pessoa, Para\'{\i}ba, Brazil}
\author{Carlos M. Reyes}
\email{creyes@ubiobio.cl}
\affiliation{Departamento de Ciencias B\'{a}sicas, Universidad del B\'{\i}o-B\'{\i}o, Casilla 447, Chill\'{a}n, Chile}
\author{Marco Schreck}
\email{marco.schreck@ufma.br}
\affiliation{Departamento de F\'{i}sica, Universidade Federal do Maranh\~{a}o, 65080-805, S\~{a}o Lu\'{i}s, Maranh\~{a}o, Brazil}
% e-mail addresses: one for each author, in the same order as the authors

\begin{abstract}
We canonically quantize $(2+1)$-dimensional electrodynamics including a higher-derivative
Chern-Simons term. The effective theory describes a standard photon and an additional degree of
freedom associated with a massive ghost. We find the Hamiltonian and the algebra satisfied by the
field operators. The theory is characterized by an indefinite metric in the Hilbert space that brings
up questions on causality and unitarity.
We study both of the latter fundamental properties and show that microcausality as well as perturbative
unitarity up to one-loop order are conserved when the Lee-Wick prescription is employed.
\end{abstract}

\maketitle

%%%%%%%%%%%%%%%%%%%%%%%%%%
\section{Introduction}
%%%%%%%%%%%%%%%%%%%%%%%%%%

The concept of an indefinite metric in a Hilbert space plays a fundamental
role in the formulation of relativistic quantum field theory.
Dirac was the first to show how
an indefinite metric arises in quantum electrodynamics and proposed how to deal with
 its probability interpretation~\cite{Dirac}.
One can mention two reasons for Dirac's suggestion.
On the one hand, any finite representation of a noncompact group --- the
Lorentz group included ---
leads to a state space endowed with an indefinite metric. On the other hand,
the commutator of two vector field operators reads
\begin{equation}
[A_{\mu}(x),A_{\nu}(y)] =\mathrm{i} \eta_{\mu \nu} D(x-y) \,,
\end{equation}
with the scalar commutator function $D$ and the Minkowski metric $\eta_{\mu\nu}$. The difference
in the signs of the metric components $\eta_{00}$ and $\eta_{ii}$ induces an indefinite metric in the corresponding
state space; see, in particular, Heisenberg's contribution in the list of references~\cite{Aspects,Heisenberg2,Pauli}.

Gupta and Bleuler used this concept within the covariant quantization of
electrodynamics. The Gupta-Bleuler formalism shows that
the unphysical
degrees of freedom are eliminated
by imposing the weak Lorentz condition on the Hilbert space.
Many of the motivations for studying indefinite metric theories come
from the theory of gravitation, where the nonrenormalizability
of the Einstein-Hilbert action forces one to consider the possibility of modified gravity theories. Some of them also
introduce indefinite metrics in the Hilbert space~\cite{Stelle,Nakanishi-QFT,Nakanishi-GR}.

The most notorious drawback of indefinite-metric theories is the possibility
of negative probabilities leading to the loss of unitarity.
Unitarity in this context has been studied extensively for the past decades.
In the sixties, Lee and Wick, being attracted by the idea
of reconciling the divergencies in quantum electrodynamics (QED)
without spoiling unitarity, constructed a modified
electrodynamics with an indefinite metric.
Their theory, which is known as Lee-Wick
model~\cite{Lee-Wick:Negative,Lee-Wick:Finite}, is a modified
electrodynamics including a massive boson field associated
with negative metric components. One characteristic of the propagator of their theory is that it contains
complex conjugate pairs of additional poles, which are called Lee-Wick poles.

The Lee-Wick model is also obtained by introducing
a higher-derivative term in the Lagrangian~\cite{Exploring}.
In this model, perturbative unitarity of the $S$-matrix
has been successfully implemented via the Cutkosky-Landshoff-Olive-Polkinghorne prescription in which
 a pair of Lee-Wick poles
cancel each other in cut diagrams~\cite{Anon-analytic}.
Several approaches have provided a deeper understanding
of many physical aspects of Lee-Wick models in the last
years~\cite{TheLee-Wick,Fior,Anewformulation,Perturbativeunitarity}.
In fact, investigations aimed at providing
finiteness in quantum field theory have not stopped, reaching
diverse application within nonlocal quantum gravity; see, e.g.,~\cite{loop,Super-Lee-Wick,Briscese}
and higher-derivative gravity extensions studied even earlier~\cite{Stelle}.

Basically, the loss of unitarity occurs due to the
negative contribution of the residue of the ghost field
to scattering cross sections. In this case
the cutting equations provided by the optical theorem cannot be satisfied.
It was demonstrated that one can modify the definition of the
internal product in the Hilbert space in order to cope with the unitarity problem.
However, this approach leads to theories characterized by non-Hermitian Hamiltonians, i.e., they
exhibit a nonstandard time evolution. However, Bender and collaborators found that
such Hamiltonians have real eigenvalues when they are symmetric under $\mathit{PT}$
transformations~\cite{Bender}. Scenarios of this kind have attracted an exceeding interest, see, e.g.,~\cite{Fring,Alexandre}
where non-Hermitian Hamiltonians are discussed, too.

Another motivation for the interest in indefinite metric theories originated from gravity where
it was demonstrated that adding higher-derivative
terms allows for gravity to be renormalizable~\cite{Stelle}.
This fact implied active studies
of renormalization of $R^2$ gravity and other higher-derivative
gravity theories (see, e.g.,~\cite{Asorey} and references therein). Nevertheless,
it was realized soon that this kind of improvement of the renormalization behavior
inevitably leads to ghosts.
From the formal viewpoint, their presence can be explained as follows.
Consider the example
of a propagator $\frac{1}{k^2(k^2+m^2)}$ occurring in a fourth-derivative theory.
A simple transformation shows that
this propagator describes two particles: a massless and
a massive one. The propagator of the latter carries a negative sign, whereby
the massive particle corresponds to a
free scalar field with possibly negative energy. Even if
the energy in the theory can be bounded from below due to a redefinition of vacuum,
unitarity, upon the presence of interactions, is expected to be broken
(see~\cite{HH,E-W} for more detailed explanations).

Furthermore, more problems related to the consistent
quantum description of higher-derivative theories were
discussed in~\cite{Ant,Anewformulation,Perturbativeunitarity}.
In the latter papers, it was claimed that these problems actually arise
due to differences between the behaviors of the theory
in Minkowski spacetime and its counterpart in Euclidean space.
At the same time, it was argued in~\cite{Smilga} that in certain
cases the ghosts are \q{benign} so that the theory turns out to be perturbatively
unitary, with the vacuum being perturbatively stable. Therefore, the
problem of ghosts must be considered separately for any
higher-derivative theory.

An interesting example of a higher-derivative
extension of quantum electrodynamics (QED) containing dimension-five operators
was proposed by Myers and Pospelov~\cite{MP}.
The higher-derivative term in its Lagrangian, called the
Myers-Pospelov term, involves explicit Lorentz symmetry breaking,
so that for some special choice of the Lorentz-breaking
preferred four-vector, higher time derivatives do not arise,
whereupon unitarity breaking is avoided. In case an indefinite metric
occurs, one can apply the Lee-Wick prescription to
show that unitarity is conserved~\cite{tree1,loop1,Unit_loop}.
According to the latter, all negative-norm states are removed
from the asymptotic Hilbert space. This procedure will turn out to be fruitful
in the analysis that we intend to carry out in the current paper.

A further interesting Lorentz-breaking modification
of QED is the higher-derivative Carroll-Field-Jackiw-like term exhibiting a
similar behavior (both of these terms were shown to be generated perturbatively at the
one-loop level, whereby the corresponding contributions are finite, see~\cite{MNP}).
In a different context, though,
the possibility of Lorentz violation
due to an indefinite metric
was pointed out several years ago by Nakanishi~\cite{LIV-LW,Nakanishi-LIV}.

Therefore, to understand the physical impact of effective
higher-derivative extensions of QED, it is important to check how
such terms affect unitarity.
To do so, though, it is reasonable to investigate a simplified model
first, that is, $(2+1)$-dimensional QED with an additive higher-derivative
Chern-Simons (CS) term, which does not involve Lorentz symmetry breaking.
Some classical issues related to this theory such as the nature and behavior of
degrees of freedom were analyzed earlier in~\cite{DJ}. Its canonical formulation was
discussed in~\cite{testing} and the perturbative
generation of the higher-derivative CS term was carried out in~\cite{Passos}.
Here, we intend to elaborate on the aspects of
microcausality and unitarity of this theory.

The structure of the paper looks as follows.
In Sec.~\ref{sec2}, we introduce the classical
action and the propagator of our theory, write down the classical field
equations, the dispersion equation, and its solutions. Furthermore, we
decompose the higher-derivative theory into a standard one involving degrees of freedom associated with
a three-component photon field and a second contribution in terms of a Proca ghost field.
We then find the polarization vectors for the photon and the massive ghost as well as their stress tensors.
In Sec.~\ref{sec3}, we canonically quantize the theory, construct the field operators such that they satisfy the expected algebra,
 and analyze the constraint structure in combination with
finding the Hamiltonian. In Sec.~\ref{sec4}, we verify tree-level unitarity of our theory and
we also study perturbative unitarity at one-loop level. Section~\ref{sec5}
states a final summary and discussion of our results. Appendix~\ref{App:A} contains
details of the derivation of Dirac brackets
and the Dirac formalism that reduces second-class constraints to zero. Appendix~\ref{App:B}
explains how to express the Hamiltonian of the theory
in terms of creation and annihilation operators. Appendix~\ref{App:C} delivers
detailed computations of the nonzero equal-time commutators satisfied by the field operators.
Finally, appendix~\ref{App:D} provides a summary of the most important properties of a Dirac
theory in $(2+1)$ dimensions.
%...............................................................................................................
\section{Higher-derivative Maxwell-Chern-Simons theory} \label{sec2}
%...............................................................................................................
In this section, we present the higher-derivative CS term coupled
to the Maxwell Lagrangian in $(2+1)$ dimensions. The theory describes
a standard photon and a massive mode at high energies associated with a ghost.
To show this, we apply a linear
transformation to the higher-derivative Lagrangian decoupling it
into a sum of two standard-derivative parts. We find the polarization
vectors and connect their sum with the propagator, which
simplifies the study of unitarity in Sec.~\ref{sec4}.
%------------------------------------------------------------------------
\subsection{The $(2+1)$-dimensional model}
%------------------------------------------------------------------------
Our starting point consists of a Lagrangian that is the sum of the
standard Maxwell term and the higher-derivative CS extension in $(2+1)$ dimensions~\cite{DJ},
 given by
\begin{equation}\label{L-D}
\mathcal{L}=-\frac{1}{4}F_{\mu\nu}F^{\mu\nu}
+\frac{g}{2}\epsilon^{\alpha\beta\gamma}(\Box
 A_\alpha)(\partial_\beta A_\gamma)+\mathcal L_{\mathrm{GF}}\,,
\end{equation}
where $\square=\partial^{\mu}\partial_{\mu}$ is the d'Alembertian
and $g$ is a small constant with inverse mass dimension. We will
see that the inverse of $g$ is related to a mass scale. Thus, it is assumed that
$g>0$. Furthermore, $\mathcal{L}_{\mathrm{GF}}$
is a covariant gauge-fixing term
inversely proportional to the arbitrary gauge-fixing parameter $\xi$
\begin{equation}\label{gauge-fixing}
\mathcal{L}_{\mathrm{GF}}=-\frac{1}{2\xi}(\partial_\mu A^{\mu})^2\,.
\end{equation}
We take the metric convention $\eta_{\mu \nu} =\text{diag} (+,-,-)$,
and our definition of the Levi-Civita symbol is based on
$\epsilon^{012}=\epsilon_{012}=\epsilon^{12}=\epsilon_{12}=1$. Hence, all Lorentz indices run over 0,1,2.

Within our study we do not consider the usual single-derivative CS term
for the sake of simplicity, since we aim at keeping track of the
higher-derivative contribution. We note that the CS term is suppressed above some energy scale
in comparison to our higher-derivative term. In principle, though,
it is natural to expect that it would not render the physics
essentially different. Nevertheless, the complete analysis of unitarity and,
especially, of the Dirac algebra of constraints will be much more involved if the CS term is present.
Therefore, we discard it in our analysis.

We note in passing that a $(2+1)$-dimensional Lorentz-violating electromagnetism involving
higher-derivative terms was derived in \cite{Ferreira:2019jbx} from the electromagnetic sector of the nonminimal
Standard-Model Extension \cite{Kostelecky:2009zp} via a procedure known as dimensional reduction (see, e.g.,
\cite{Belich:2002vd,Belich:2003xa}). The second contribution
in Eq.~(\ref{L-D}) can be mapped onto the third one in $\mathcal{L}_{(1+2)}$ of \cite{Ferreira:2019jbx} via suitable
partial integrations.

The treatment of systems in classical mechanics described by
higher-derivative Lagrangians was initiated by Ostrogradsky
in his seminal paper~\cite{Ostro}. Subsequent scientific papers reviewing and extending his original
ideas are~\cite{Borneas:1959,Riahi:1972,Woodard:2015} where this list is not claimed
to be exhaustive. One of the central
results of these works is that an application of the Hamilton principle leads to a modified set
of Euler-Lagrange equations. An analogous development of the formalism in the
context of higher-derivative field theory can be found, e.g., in~\cite{Bollini:1986am}.
For the particular field theory defined by Eq.~(\ref{L-D}), it is sufficient to restrict
these generalized Euler-Lagrange equations to
\begin{equation}
-\partial_{\kappa} \partial_{\lambda}  \frac{\partial \mathcal L}
{\partial( \partial_{\kappa} \partial_{\lambda}  A_{\sigma})  }
+ \partial_{\rho}  \frac{\partial \mathcal L}
{\partial( \partial_{\rho}   A_{\sigma})  }-   \frac{\partial \mathcal L}
{\partial  A_{\sigma}  }=0\,.
\end{equation}
They lead to the modified Maxwell equations
\begin{equation}\label{E-L1}
\partial _{\rho}F^{\rho \sigma}+g  \epsilon^{\sigma\beta\gamma} \Box \partial_\beta {A}_\gamma
+\frac{1}{\xi} \partial^{\sigma}(\partial \cdot A)=0\,.
\end{equation}
Now, contracting Eq.~(\ref{E-L1}) with $\partial_{\sigma}$ yields
\begin{equation}
\frac{1}{\xi}\Box \left(\partial \cdot A \right)=0\,.
\end{equation}
Hence, by imposing suitable boundary conditions at infinity it follows that
$\partial \cdot A=0$ can be set.

Now, let us rewrite the Lagrangian~(\ref{L-D}) as
\begin{equation}\label{eq_mot_der}
\mathcal{L}=\frac{1}{2}A_\mu\left[ \Box \eta^{\mu \nu}
-\left(1-\frac{1}{\xi}\right)\partial^\mu\partial^\nu
+g\epsilon^{\mu\beta\nu}   \partial_\beta\Box \right]  A_\nu\,,
\end{equation}
yielding the equations of motion for the gauge field:
\begin{equation}\label{G-F-E}
\left[\Box  \eta^{\mu \nu}
-\left(1-\frac{1}{\xi}\right)\partial^\mu \partial^\nu
+g\epsilon^{\mu \beta\nu}   \partial_\beta \Box   \right]  A_\nu (x) =0\,.
\end{equation}
Transforming the latter to the
 momentum representation with $\mathrm{i}\partial_{\mu}=p_{\mu}$, we write
\begin{subequations}
\begin{equation}\label{eq_S}
S^{\mu \nu}(p) A_{  \nu}(p)=0   \,,
\end{equation}
with
\begin{equation}\label{S-eq}
S^{\mu \nu}(p)=p^2\left[\eta^{\mu\nu}-\left(1-\frac{1}{\xi}\right)\frac{ p^{\mu}p^{\nu}}{p^2}
 -\mathrm{i}g \epsilon^{\mu\beta\nu}p_\beta \right]  \,.
\end{equation}
\end{subequations}
The propagator $P_{\mu\nu}$ follows from inverting the operator
$S^{\mu\nu}$, giving
\begin{subequations}
\label{eq:propagator-pmunu}
\begin{equation}
\label{prop_xi}
P_{\mu\nu}(p)=-\frac{G_{\mu \nu}(\xi,p)}{p^2 (1-g^2p^2)} \,,
\end{equation}
where
\begin{equation}\label{prop_G}
G_{\mu\nu}(\xi,p)=   \eta_{\mu \nu}-
  \left[1-\xi \left(1-g^2 p^2\right)\right]   \frac{p_{\mu} p_{\nu}}{p^2}+\mathrm{i}g\epsilon_{\mu
\beta \nu}p^{\beta} \,.
\end{equation}
\end{subequations}
The conventions have been chosen such that the propagator satisfies
\begin{equation}
S^{\mu \nu}(p)P_{\nu \rho}(p)=-\delta^{\mu}_{\phantom{\mu}\rho}\,.
\end{equation}
Considering the pole structure of the propagator~(\ref{eq:propagator-pmunu}) and defining
$g\equiv M^{-1}$, we decompose the denominator as
 \begin{equation}\label{descp}
\frac{  M^2 }{p^2(p^2- M^2)}=-\frac{1}{p^2}+\frac{1}{p^2-M^2}\,,
\end{equation}
where the second contribution has a residue whose sign is opposite to that
of the first contribution. Hence, it can be associated
with a ghost. The dispersion relations are given by the propagator poles with
respect to $p_0$. Determining the poles yields the modes
corresponding to a photon and a massive gauge field given by
\begin{subequations}
\begin{eqnarray} \label{Sol1}
\omega(\vec p)&=\omega_p=&|\vec p|\,,
\\[2ex] \label{Sol2}
  \Omega (\vec p)&=\Omega_p=&\sqrt{\vec p^{\,2}+ M^2}\,,
\end{eqnarray}
\end{subequations}
respectively.

Let us write down the energy-momentum tensor of our theory.
It is clear that it is a sum of two contributions. The first is the
energy-momentum tensor for electrodynamics in $(2+1)$ dimensions
whose symmetric form is the well-known Belinfante tensor equal to
\bea
T_{\mathrm{Bel}}^{\mu\nu}=F^{\mu}_{\phantom{\mu}\lambda}F^{\lambda\nu}
+\frac{1}{4}\eta^{\mu\nu}F_{\alpha\beta}F^{\alpha\beta}\,.
\eea
The second is connected to the higher-derivative Chern-Simons (HDCS) theory, whose symmetric
 form was found explicitly in~\cite{DJ}. So we merely quote the result, which is
\begin{equation} \label{eq:stress-tensor-hdcs}
T^{\mu\nu}_{\mathrm{HDCS}}=g [(\epsilon^{\mu\alpha\beta}F^{\ast \nu}
+\epsilon^{\nu\alpha\beta}F^{\ast \mu}) \partial_{\alpha}
F^{\ast}_{\beta}-\eta^{\mu\nu}\epsilon^{\alpha\beta\gamma}
F^{\ast}_{\alpha}\partial_{\beta}F^{\ast}_{\gamma}
],
\end{equation}
where $F^{\ast}_{\alpha}=\frac{1}{2}\epsilon_{\alpha\mu\nu}
F^{\mu\nu}$ is the dual of the field strength tensor $F^{\mu\nu}$.
%..........................................................
\subsection{Decoupling the ghost}
%..........................................................
Here we make explicit the two types of fields described by the Lagrangian~(\ref{L-D}).
We define the new fields as
\begin{subequations}
\label{eq:collection-new-fields}
\begin{eqnarray}
\label{new_fields}
{\bar A}_{\mu}&=&\frac{1}{\sqrt{2}}(A_\mu+gF^{\ast}_{\mu})\,, \\[2ex]
{G}_\mu&=&\frac{g}{\sqrt{2}}F^{\ast}_{\mu}\,, \label{new_fields2}
\end{eqnarray}
\end{subequations}
in terms of the dual tensor $F_{\mu}^{\ast}$ defined under Eq.~(\ref{eq:stress-tensor-hdcs}) and the
original photon field $A_\mu$.

Considering Eqs.~(\ref{new_fields}) and (\ref{new_fields2}), we find the identities
\begin{subequations}
\begin{align}
-\frac{1}{4}\bar F_{\mu\nu}\bar F^{\mu\nu}&= -\frac{1}{8} F_{\mu\nu}
F^{\mu\nu} \notag \\
&\phantom{{}={}}-\frac{g}{2}\left(\partial_{\mu}A_{\nu}   +\frac{g}{2}  \partial_{\mu} F^{\ast}_{\nu}  \right)
(\partial^{\mu} F^{\ast \nu} -\partial^{\nu} F^{\ast \mu}    )\,,
\end{align}
and
\begin{eqnarray}
-\frac{1}{4}  F^{\ast}_{\mu} F^{\ast \mu} = -\frac{1}{8} F_{\mu\nu} F^{\mu\nu} \,,
\end{eqnarray}
\end{subequations}
where $\bar F_{\mu \nu}= \partial_{\mu}\bar A_{\nu}- \partial_{\nu}\bar A_{\mu} $ is the field strength
tensor associated with the new field of Eq.~(\ref{new_fields}).

Now, by adding both equations, performing suitable integrations by parts, and using
the (unmodified) homogeneous Maxwell equation $\partial_{\mu}F^{*\mu}=0$ in $(2+1)$ dimensions, we
can rewrite the first part of the Lagrangian~(\ref{L-D}) as
\begin{align}
&-\frac{1}{4}F_{\mu\nu} F^{\mu\nu}+\frac{g}{2}
\epsilon^{\alpha \beta \gamma}(\Box A_\alpha)(\partial_\beta A_\gamma)
\nonumber \\ &=-\frac{1}{4}\bar{F}_{\mu\nu}\bar{F}^{\mu\nu}-\frac{g^2}{4}
F^{\ast}_{\mu}    \left(\frac{1}{g^2}+\Box\right) F^{\ast \mu} \,.
\end{align}
Using the definition~(\ref{new_fields2}) and $\partial_{\mu} { A^ \mu}=\sqrt 2\, \partial_{\mu} {\bar A ^\mu}$
allows us to write the higher-derivative Lagrangian as the sum
\begin{align}
\mathcal{L}&=-\frac{1}{4}\bar{F}_{\mu\nu}\bar{F}^{\mu\nu}-\frac{1}{\xi}(\partial_\mu\bar{A}^\mu)^2 \notag \\
&\phantom{{}={}}+\frac{1}{2} \partial_{\mu}{G }_\nu  \partial^{\mu}{G }^\nu-\frac{1}{2}M^2  {G }_\mu{G }^\mu   \,,
\end{align}
where the higher derivatives have been absorbed into the new fields.
The first part of the new Lagrange density describes a photon with a gauge fixing term
and the second part corresponds to a Proca field theory involving a mass
scale of the order of $M\sim g^{-1}$. As the coupling constant $g$ of the
modification is assumed to be small, the latter mass scale $M$ is supposed to be large.
The Proca field theory presumably describes a ghost dominating the regime of high energies.
%..........................................................................................
\subsection{Polarization vectors}
%.........................................................................................
Now that the theory has been decomposed into two decoupled
standard-derivative parts associated with the fields of Eqs.~\eqref{new_fields} and \eqref{new_fields2},
our next step is to find the polarization vectors.
First, they are crucial for the computation of the Hamiltonian in
terms of creation and annihilation operators. Second, they are
needed to construct the tensor structure in the equal-time commutation relations
of the field operators. Last but not least, the propagator can be expressed
in terms of the polarization vectors, which will be helpful to prove the validity
of the optical theorem.

To begin with, consider the following orthogonal basis of $(2+1)$-dimensional
Minkowski spacetime that involves the three real vectors
\begin{subequations}
\label{eq:pol-all}
\begin{align}\label{eq:pol0}
e^{(0)\mu}&=\frac{1}{\sqrt{p^2}} p^\mu,
\displaybreak[0]\\
\label{eq:pol1}
e^{(1)\mu}&=\frac{1}{\sqrt{G}} \epsilon^{\mu\beta\gamma}p_\beta n_\gamma,
\displaybreak[0]\\
\label{eq:pol2}
e^{(2)\mu}&=-\frac{1}{\sqrt{p^2}} \epsilon^{\mu\beta\gamma}p_\beta e^{(1)}_\gamma
=\frac{1}{\sqrt{ p^2 G}} (p^2n^{\mu}- p^{\mu} (p\cdot n)),
\end{align}
\end{subequations}
where $G=(p\cdot n)^2-p^2n^2$ and $n^{\mu}$ is an auxiliary three-vector.
The three-vectors $e^{(a)}_{\mu}$ are normalized according to
\begin{equation}\label{normalization}
e^{(a)}  \cdot  e^{(b)}=g_{ab}\,,
\end{equation}
with $a=0,1,2$ and $ g_{ab} =\text{diag} (1,-1,-1)$.
Although $g_{ab}$ formally corresponds to the Minkowski metric in $(2+1)$
dimensions, we use another symbol here, as the indices of this object are not Lorentz indices,
but merely the labels of the vectors introduced before.
In order to ensure $G>0$, we will take $p^2>0$ and choose $n^{\mu}$
as a timelike vector.

Furthermore, these vectors satisfy the completeness relation
\begin{equation}
\sum_{a,b=0} ^2 g_{ab}\, e^{(a)}_\mu e^{(b)}_\nu=\eta_{\mu\nu}\,.
\end{equation}
However, note that the above basis is not suitable to describe the photon field
due to the denominator depending on $\sqrt{p^2}$. To construct suitable polarization
vectors for photons
we will proceed differently in the subsection~\ref{Ham0}.

Moreover, one can check that the $e^{(a)}_{\mu}$ fulfill the relations
\begin{subequations}
\begin{eqnarray}\label{curl}
\epsilon^{\mu\beta\gamma}p_\beta e^{(2)}
_\gamma&=& \sqrt{p^2} e^{(1)\mu}   \,, \\[2ex]
\epsilon^{\mu\beta\gamma}p_\beta e^{(1)}
_\gamma&=&- \sqrt{p^2} e^{(2)\mu} \,.
\end{eqnarray}
\end{subequations}
With the real basis $\{e^{(a)}\}$ at hand, we look
for a complex basis $\{\varepsilon^{(\lambda)}\}$
diagonalizing the operator $S^{\mu}_{\phantom{\mu}\nu}(p)$ of
Eq.~(\ref{S-eq}). Our intention is to relate the propagator to the sum of polarization
tensors formed from the vectors of $\{\varepsilon^{(\lambda)}\}$. This particular
method was introduced in~\cite{CMP} and applied in the context of the Maxwell-Chern-Simons-like
theory in $(3+1)$ dimensions. We adopt it to the theory of Eq.~(\ref{L-D}), as it turned
out to be valuable for checking the validity of the optical theorem.
Hence, considering Eq.~(\ref{S-eq}) we demand
that these vectors fulfill
\begin{equation}
S^{\mu}_{\phantom{\mu}\nu}(p) \varepsilon^{(\lambda)   \nu}(p)=\Lambda_{\lambda}(p)
  \varepsilon^{(\lambda) \mu}(p)   \,,
\end{equation}
with the new label $\lambda\in \{0,+,-\}$ and the eigenvalue
 $\Lambda_{\lambda}(p)$ of the polarization mode $\lambda$.

We now define the complex basis as follows:
\begin{subequations}
\label{complex_basis}
\begin{align}
\varepsilon^{(0)\mu}&=e^{(0)\mu}   \,, \displaybreak[0]\\[2ex]
\varepsilon^{(+)\mu}&=\frac{e^{(2)\mu}+\mathrm{i}e^{(1)\mu}}{\sqrt{2}}\,, \displaybreak[0]\\[2ex]
\varepsilon^{(-)\mu}&=\frac{e^{(2)\mu}-\mathrm{i}e^{(1)\mu}}{\sqrt{2}}\,.
\end{align}
\end{subequations}
The $\pm$ modes are orthogonal to the momentum, that is,
$p\cdot \varepsilon^{(\pm)}=0$. By using~Eqs.~(\ref{normalization})
and~(\ref{curl}) one can show that
\begin{subequations}
\label{eq:relations-epsilon}
\begin{eqnarray}
\varepsilon^{(\lambda)} \cdot \varepsilon^{(\lambda')*} &=&g_{\lambda\lambda'  }\,, \\[2ex]
\epsilon^{\mu\beta\sigma}p_\beta \varepsilon^{(\pm)}
_\sigma&=&\mp \mathrm{i} \sqrt{p^2} \varepsilon^{(\pm)\mu}  \,,
\end{eqnarray}
\end{subequations}
with $g_{\lambda \lambda'}=\text{diag}(1,-1,-1)$. Note that the latter
matrix again corresponds to the Minkowski metric in $(2+1)$ dimensions. As its indices
are the labels of the vectors $\{\varepsilon^{(\lambda)}\}$, we denote it by $g_{\lambda\lambda'}$.

Indeed, it is not difficult to show that the vectors of Eq.~(\ref{complex_basis})
diagonalize $S^{\mu\nu}$, i.e.,
\begin{subequations}
\begin{eqnarray}
S^\mu_{\phantom{\mu}\nu}(p)\varepsilon^{(0)\nu}&=&
\Lambda_{0}(p)\varepsilon^{(0)\mu}  \,, \\[2ex]
S^\mu_{\phantom{\mu}\nu}(p)\varepsilon^{(+)\nu}&=&
\Lambda_{+}(p)\varepsilon^{(+)\mu}    \,, \\[2ex]
S^\mu_{\phantom{\mu}\nu}(p)\varepsilon^{(-)\nu}&=&\Lambda_{-}(p)\varepsilon^{(-)\mu}\,,
\end{eqnarray}
\end{subequations}
where the eigenvalues are given by
\begin{subequations}
\label{eigen_L}
\begin{eqnarray}
\Lambda_{0}(p)&=&\frac{p^2}{\xi}\,,   \label{eigen_L1} \\[2ex]
\Lambda_{+}(p)&=&p^2\left(1-g\sqrt{p^2}\right) \,,\label{eigen_L2}  \\[2ex]
\Lambda_{-}(p)&=&p^2\left(1+g\sqrt{p^2}\right)\,. \label{eigen_L3}
\end{eqnarray}
\end{subequations}
The dispersion relations of our theory follow from requiring that
the product of eigenvalues vanish,
\begin{equation}\label{disp-relation}
\prod_{\lambda=0,\pm} \Lambda_{\lambda}(p)
=\frac 1 \xi(p^2)^3 (1-g^2p^2)=0\,,
\end{equation}
giving the dispersion relations of Eqs.~(\ref{Sol1}) and (\ref{Sol2})
for the photon and massive ghost mode, respectively. Hence, the vectors of the basis
$\{\varepsilon^{(\lambda)}\}$ are solutions of the field equations when they are
evaluated on-shell. Therefore, they can be interpreted as polarization vectors.

From these relations, it is possible to show that
\begin{eqnarray}\label{sum_T}
\varepsilon^{(\pm)}_\mu \varepsilon^{(\pm)*}_\nu=-
\frac{1}{2}\left(\eta_{\mu\nu}-\frac{p_\mu p_\nu}
{p^2}\pm \mathrm{i} \frac{\epsilon_{\mu\beta\nu}p^\beta}{\sqrt{p^2}}\right)\,,
\end{eqnarray}
or
\begin{eqnarray}
\label{eq:polarization-tensor-epsilon}
\varepsilon^{(\pm)}_\mu \varepsilon^{(\pm)*}_\nu=
\frac{1}{2}\left(  e_{\mu \nu}\pm\mathrm{i}\epsilon_{\mu \nu}  \right)\,,
\end{eqnarray}
where we have defined the tensors $e_{\mu \nu}$ and $\epsilon_{\mu \nu}$
by
\begin{subequations}
\begin{eqnarray}
e_{\mu \nu}\equiv e^{(1)}_\mu   e^{(1)}_\nu +e^{(2)}_\mu   e^{(2)}_\nu
&=&-\eta_{\mu \nu}+\frac{p_{\mu} p_{\nu}   }{p^2}\,, \\[2ex]
\epsilon_{\mu \nu}\equiv e^{(1)}_\mu   e^{(2)}_\nu -e^{(1)}_\nu   e^{(2)}_\mu
&=&-\frac{1  }{\sqrt{p^2}}\epsilon_{\mu \beta \nu }   p^{\beta} \,.
\end{eqnarray}
\end{subequations}
Now, to make contact with the propagator $P_{\mu\nu}$ of Eq.~(\ref{prop_xi}) via the relation~\cite{CMP}
\begin{equation} \label{sum_propagator}
P_{\mu\nu}=-\sum_{\lambda,\lambda'=0,\pm}g_{\lambda\lambda'}\frac{
\varepsilon^{(\lambda)}_\mu \varepsilon^{(\lambda')*}_\nu}{\Lambda_{\lambda}}   \,,
\end{equation}
we consider the sum over two-tensors formed from the polarization vectors.

First, we investigate the transverse part and perform the sum over the $\pm$ modes.
Based on the eigenvalues of Eqs.~(\ref{eigen_L1}), (\ref{eigen_L2}), (\ref{eigen_L3}) and the finding
of Eq.~(\ref{sum_T}), we have
\begin{align}
\frac{\varepsilon ^{(+)}_\mu \varepsilon ^{(-)}_\nu}{\Lambda_+}&+
\frac{\varepsilon ^{(-)}_\mu \varepsilon ^{(+)}_\nu}{\Lambda_-}
\nonumber \\
&\hspace{-0.8cm}=-\frac{1}{p^2(1-g^2p^2)}\left(\eta_{\mu\nu}-\frac{p_\mu p_\nu}{p^2}+
\mathrm{i}g\epsilon_{\mu\beta\nu}p^\beta \right)\,.
\end{align}
Next, by adding the mode labeled with $\lambda=0$ we obtain
\begin{align}
\frac{\varepsilon^{(0)}_\mu \varepsilon ^{(0)}_\nu}{\Lambda_0}&
-\frac{\varepsilon^{(+)}_\mu \varepsilon ^{(-)}_\nu}
{\Lambda_+}-\frac{\varepsilon ^{(-)}_\mu \varepsilon ^{(+)}_\nu}{\Lambda_-} \nonumber \\
&\hspace{-0.8cm}=\frac{1}{p^2(1-g^2p^2)}\left(\eta_{\mu\nu}-\frac{p_\mu p_\nu}{p^2}+\mathrm{i}g
\epsilon_{\mu\beta\nu}p^\beta \right) +\frac{\xi p_\mu p_\nu}{(p^2)^2}\,, \nonumber \\
\end{align}
to finally arrive at
\begin{align}
&-\sum_{\lambda,\lambda'=0,\pm}g_{\lambda\lambda'}\frac{
\varepsilon^{(\lambda)}_\mu \varepsilon^{(\lambda')*}_\nu}{
\Lambda_{\lambda}}\nonumber  \\
&=-\frac{1}{p^2 (1-g^2p^2)} \nonumber \\
&\phantom{{}={}}\times\left[\eta_{\mu \nu}-
  \left(  1-\xi \left(1-g^2 p^2\right)\right)   \frac{p_{\mu} p_{\nu}}{p^2}+\mathrm{i}g\epsilon_{\mu
\beta \nu}p^{\beta}\right]\,.
\end{align}
The latter is just the propagator of~Eq.~(\ref{eq:propagator-pmunu}). Hence,
the method introduced in~\cite{CMP,Don1} turns out to work in the context of the
$(2+1)$-dimensional theory defined by Eq.~(\ref{L-D}), as well.
%%%%%%%%%%%%%%%%%%%%%%%%%%
\section{Canonical quantization} \label{sec3}
%%%%%%%%%%%%%%%%%%%%%%%%%%
In this section, we quantize the higher-derivative theory starting from the extended symplectic
structure provided by the Ostrogradsky formalism~\cite{Ostro,Borneas:1959,Riahi:1972,Woodard:2015}
applied to the context of higher-derivative field theories~\cite{Bollini:1986am}.
The theory of Eq.~(\ref{L-D}) has constraints
that modify the canonical Poisson brackets rendering its quantization more involved.
We compute the Hamiltonian by choosing a particular vacuum state
and show that the theory
is stable, but the associated Hilbert space is endowed with an indefinite metric.
We prove that in spite of the presence of negative-norm states, which can be interpreted as ghosts,
causality is preserved in the theory.
%..........................................................................................
\subsection{Constrained Hamiltonian formulation} \label{Ham0}
%..........................................................................................
We consider the Lagrangian~(\ref{L-D}) for $\xi=1$ and
after some integration by parts we arrive at
\begin{equation}\label{Lag_CS}
\mathcal{L}=-\frac{1}{2} \partial_{\mu}A_\nu \partial^{\mu}   A^\nu
+\frac{1}{2}g\epsilon^{\mu\beta\gamma}   \Box   A_\mu    \partial_\beta A_\gamma    \,.
\end{equation}
The variational methods of higher-derivative
theories~\cite{Ostro,Borneas:1959,Riahi:1972,Woodard:2015,Bollini:1986am} are applied
to obtain the canonical conjugated momenta to both $A_{\mu}$ and $\dot A_{\mu}$.
They are given by
\begin{subequations}
\begin{eqnarray}
P^\mu&=&\frac{\partial \mathcal{L}}{\partial\dot{A}_\mu}-\frac{\partial\Pi^\mu}{\partial t}  \,,
\\[2ex]
\Pi^\mu&=&\frac{\partial\mathcal{L}}{\partial\ddot{A}_\mu}\,,
\end{eqnarray}
\end{subequations}
respectively.
The higher-order Hamiltonian follows from an extended Legendre transformation,
\begin{equation}\label{higher-Ham}
H=\int \mathrm{d}^2x\left(P^\mu(x)\dot{A}_\mu(x)+\Pi^\mu(x)\ddot{A}_\mu(x)-\mathcal L(x)\right)\,,
\end{equation}
and the canonical Poisson brackets for the extended phase space are
\begin{subequations}
\begin{eqnarray}
\{A_{\mu}(t,\vec x), P_{\nu}(t,\vec y) \} &=&\eta_{\mu \nu}\delta^{(2)}(\vec x-\vec y) \,, \\[2ex]
\{ \dot A_{\mu}(t,\vec x), \Pi_{\nu}(t,\vec y) \} &=&\eta_{\mu \nu} \delta^{(2)}(\vec x-\vec y) \,,
\end{eqnarray}
\end{subequations}
where the remaining ones vanish.

Applying these formulas to the specific Lagrangian~(\ref{Lag_CS}) one finds
\begin{subequations}
\begin{eqnarray}\label{second-cons}
P^\mu&=&-\dot A^{\mu}-\frac{g}{2}\epsilon^{\mu0\gamma}\Box A_\gamma- \frac{g}{2}
\epsilon^{\mu\beta\gamma}\partial_\beta  \dot A_\gamma  \,, \\[2ex]  \label{pi}
\Pi^\mu&=&    \frac{g}{2}  \epsilon^{\mu\beta\gamma}\partial_\beta  A_\gamma\,.
\end{eqnarray}
\end{subequations}
After inserting them into~Eq.~(\ref{higher-Ham}), the Hamiltonian reads
\begin{subequations}
\begin{align}
\label{Ham}
H&=\int \mathrm{d}^2x\,\bigg( - \frac 12 \dot{A}_\mu   \hat U ^{\mu \nu } \dot{A}_\nu + \frac{1}{2} A_\mu
 \hat U ^{\mu \nu }   \nabla^2 A_\nu \notag \\
&\phantom{{}={}}\hspace{1.3cm}+\frac{g}{2}\epsilon^{ij }  \dot{A}_i \Box A_j\bigg)\,,
\end{align}
where we have defined the tensor operator
\begin{equation}\label{def_U}
\hat U^{\mu \nu }=\hat{U}^{\mu\nu}(\partial)=\eta^{\mu \nu}+g\epsilon^{\mu \beta \nu} \partial_{\beta}\,.
\end{equation}
\end{subequations}
Recall the Levi-Civita symbol in $(2+1)$ dimensions defined below Eq.~(\ref{gauge-fixing}).

In order to quantize the theory, as usual, one postulates equal-time
commutation relations on the phase space variables:
\begin{subequations}
\begin{eqnarray}
\left[   A _{\mu}(t,\vec x) ,P_\nu(t,\vec y)
 \right]&=&\mathrm{i} \eta_{\mu \nu} \delta^{(2)}(\vec x-\vec y)  \,,
\\[2ex] \label{comm2}
\left[   \dot A_{\mu} (t,\vec x) ,\Pi_{\nu}(t,\vec y)
 \right]&=&\mathrm{i} \eta_{\mu \nu} \delta^{(2)}(\vec x-\vec y)  \,,
\end{eqnarray}
\end{subequations}
where all the others are defined to vanish.

However, for constrained systems, the above commutators are not always possible
to be satisfied~\cite{Constrained-Systems}. For instance, taking
the derivative $ \frac{g}{2}  \epsilon^{\mu\beta\gamma}\partial_\beta$ of the
first field of the commutator
\begin{equation}
\left[    A _{\mu}(t,\vec x) ,\dot A_\nu(t,\vec y) \right]=0\,,
\end{equation}
producing $\Pi(t,x)$, gives a relation incompatible with the commutator of~Eq.~(\ref{comm2}).
Therefore, the canonical structure of constraints has to be taken into consideration
in order to modify the Poisson brackets consistently. Some work in this direction
has already been carried out; see the formulation of first- and second-class
constraints for the higher-derivative Maxwell-Chern-Simons theory in~\cite{testing,Mukherjee,Sararu}.
In the latter papers, the Dirac approach has been implemented and the
reduced Hamiltonian has been obtained successfully with
second-class constraints strongly imposed to zero. The Dirac brackets
together with the reduced Hamiltonian neatly reproduce the equations of motion.

Here, in order to implement quantization we will follow an alternative method.
We will quantize the fields such that they satisfy the second-class constraints
automatically via their expansion in terms of plane waves.
That is, in addition to requiring that the plane waves propagate with energy $\omega_p$
of Eq.~(\ref{Sol1}) and $\Omega_p$ of~Eq.~(\ref{Sol2}), respectively, we choose the
polarization vectors such that the fields satisfy the equations of motion and the
second-class constraints in the Dirac formalism; see Appendix~\ref{App:A}.
Then, we expect the fields $A_{\mu}(t,\vec x)$ and $\dot A_{\mu}(t,\vec x)$ together
with their canonical conjugate momenta to reproduce the Dirac algebra.
We verify this property for each relevant field operator in Appendix~\ref{App:C}.
Notice, though, that the field $A_{\mu}(t,\vec x)$ cannot be considered
physical in the sense of propagating degrees of
freedom independent of the gauge fixing
parameter $\xi$. In Lorenz gauge, there is still the unphysical polarization
vector associated with the mode $\lambda=0$.

Let us consider the decomposition of our gauge field $A_{\mu}$ in terms of the
photon and massive ghost field of Eqs.~(\ref{new_fields}), (\ref{new_fields2}) as follows:
\begin{equation} \label{planewave}
 A_\mu(x)=\bar  A_\mu(x)+G_\mu(x)\,.
\end{equation}
By inserting the decomposition into the equation of motion~(\ref{G-F-E}) with
$\xi=1$ and considering the on-shell condition for the photon, $\Box \bar A_{\mu}=0$, we arrive at
\begin{equation}\label{constraints-second}
\left(\eta^{\mu \nu}+g\epsilon^{\mu \beta\nu}   \partial_\beta    \right)  G_{\nu}=0\,.
\end{equation}
By taking the derivative $\partial_{\mu}$ of~Eq.~(\ref{constraints-second}), one has
\begin{equation}\label{ghost-ort}
\partial \cdot  G=0\,.
\end{equation}
Considering all these conditions, we can write the photon field operator as
\begin{align}
\label{photon-field}
\bar A_\mu(x)&=\int\frac{\mathrm{d}^2 \vec p}{(2\pi)^2}\sum_{\lambda=0,1,2} \frac{1}{2\omega_p}  \Big[
 a^{( \lambda)}_{\vec{p}}\,  \bar{e}^{(\lambda)}_{\mu}   (  p)\, \mathrm{e}^{-\mathrm{i} p \cdot x}  \nonumber \\
&\phantom{{}={}}\hspace{1.5cm}+a^{(\lambda )\dag}_{\vec{p}} \, \bar {e}^{(\lambda)*}_{\mu}  ( p)\,
\mathrm{e}^{\mathrm{i}p\cdot x}\Big]  _{p_0=\omega_p}  \,,
\end{align}
with suitable annihilation and creation operators $a^{(\lambda)}_{\vec{p}}=a^{(\lambda)}(\vec{p})$
and $a^{(\lambda)\dag}_{\vec{p}}=a^{(\lambda)\dag}(\vec{p})$,
respectively, for the mode $\lambda$. The polarization vectors are chosen as
\begin{subequations}
 \begin{equation}
 \label{eq:polarization-vectors-e-bar}
  \bar{e}^{(\lambda)}_{\mu}   (  p) =\left(\eta_{\mu\nu}-\frac{3g^2}{8}p_\mu
p_\nu+\frac{\mathrm{i}g}{2}\epsilon_{\mu\beta\nu}p^\beta \right) \bigg|_{p_0=\omega_p}  \,
   {v}^{(\lambda)\nu}   (  p)  \,,
\end{equation}
where
\begin{align}
\label{pol_photons}
 v^{(0)\mu}(  p) &=n^\mu    \,, \displaybreak[0]\\[2ex]
v^{(1)\mu}(  p)&=\frac{ \epsilon^{\mu\beta\gamma} p_\beta
 n_\gamma}{(p\cdot n)} \bigg|_{p_0=\omega_p}   \,, \displaybreak[0]\\[2ex]
v^{(2)\mu}(  p)&=\epsilon ^ {\mu \beta \gamma}
n_{\beta}  v^{(1)}_{\gamma}(p)=\frac{p^{\mu}-n^{\mu}(p\cdot n)}{(p\cdot n)}\bigg|_{p_0=\omega_p}   \,,
\end{align}
\end{subequations}
with a timelike auxiliary vector $n_{\mu}$. Note the bar on top of the symbol
in Eq.~(\ref{eq:polarization-vectors-e-bar}) to distinguish
these vectors from the basis $\{e_{\mu}^{(a)}\}$ introduced in Eqs.~(\ref{eq:pol-all}).
One can check that the latter form an orthonormal basis, i.e.
\begin{equation}\label{u-pol}
v^{(\lambda)}_{\mu}
 v^{(\lambda')\mu}   =g^{\lambda \lambda'}\,.
\end{equation}
Also, they satisfy the relation
\begin{subequations}
\begin{equation}\label{tens-T}
\sum _{\lambda, \lambda'}  g_{\lambda \lambda'} \bar{e}^{(\lambda)}_{\mu}
  (  p)   \bar{e}^{(\lambda')*}_{\nu}   (  p)=T_{\mu \nu}(p) |_{p_0=\omega_p}\,,
\end{equation}
where we defined
\begin{align}
\label{eq:tensor-T}
T_{\mu\nu}(p)&\equiv G_{\mu \nu}(\xi=1,p) \notag \\
&=\eta_{\mu\nu}-g^2p_\mu p_\nu+\mathrm{i}g\epsilon_{\mu\beta\nu}p^\beta \,,
\end{align}
\end{subequations}
using Eq.~(\ref{prop_G}). According to~Eq.~(\ref{constraints-second})
and the orthogonality condition of~Eq.~(\ref{ghost-ort}), we write
the ghost field operator as
\begin{align}
\label{ghost-field}
G_\mu(x)&=\int\frac{\mathrm{d}^2\vec p}{(2\pi)^2}\frac{1}{2\Omega_p}
 \Big[ b_{\vec{p}}\, \bar \varepsilon_\mu^{(+)}(\vec  p)
 \, \mathrm{e}^{-\mathrm{i}p\cdot x}  \nonumber \\
&\phantom{{}={}}\hspace{2.2cm}+b^{\dag}_{\vec{p}}\,
\bar \varepsilon_\mu^{(+)*}( \vec p) \,\mathrm{e}^{\mathrm{i}p\cdot x}\Big]_  {p_0= \Omega_p}\,,
\end{align}
with another set of annihilation and creation operators $b_{\vec{p}}=b(\vec{p})$ and
$b^{\dagger}_{\vec{p}}=b^{\dagger}(\vec{p})$, respectively.
Furthermore, we defined the polarization vector
$\bar \varepsilon_\mu^{(+)}=\sqrt{2} \varepsilon_\mu^{(+)}$ in terms of the one
introduced in~Eq.~(\ref{complex_basis}). It may be convenient to make use of the property
\begin{equation}\label{tens-T-ghost}
\bar \varepsilon_\mu^{(+)}\bar \varepsilon_\nu^{(+)*}  =-T_{\mu\nu}(p)|_{p_0=\Omega_p}\,,
\end{equation}
which is equivalent to~Eq.~(\ref{eq:polarization-tensor-epsilon}). The relation $p^2=g^{-2}$
was employed to arrive at the latter result.
We impose the following algebra on the annihilation and creation
operators for the photon and ghost field:
\begin{subequations}
\label{eq:algebra-photon-ghost}
\begin{eqnarray}\label{algebra}
\left[a^{(\lambda)}_{\vec p},a^{ ( \lambda')\dag} _{\vec k}
 \right]&=&-(2\pi)^2g_{\lambda\lambda'}
2\omega_p \delta^{(2)}(\vec p-\vec k)  \,,
 \\[2ex] \label{algebra-opp}
 \left[ \, b_{\vec p} \,, \,b^{ \dag} _{\vec k }\, \right]
  &=& -(2\pi)^2
2\Omega_p \delta^{(2)}(\vec p-\vec k) \,.
\end{eqnarray}
\end{subequations}
Replacing the fields in~Eq.~(\ref{Ham}) by the field operators of
Eqs.~(\ref{photon-field}), (\ref{ghost-field}) and
using the algebra of Eqs.~(\ref{algebra}),
(\ref{algebra-opp}) and the properties of the
polarization vectors, we find the following Hamiltonian:
\begin{align}
\label{Ham2}
H&=- \frac{1}{4}\int\frac{\mathrm{d}^2\vec{p}}{(2\pi)^2}\,\Big[\sum_{\lambda,\lambda'}
 g_{\lambda\lambda'}
 \left(  a_{\vec p}  ^{(\lambda)}a^{(\lambda')\dag}_{\vec p}+
 a^{(\lambda) \dag}_{\vec p}a^{(\lambda')}_{\vec p} \right)   \nonumber \\
&\phantom{{}={}}\hspace{2.2cm}+\left(  b_{\vec p}\, b^\dag_{\vec p}+b^{\dag}_{\vec p}\, b_{\vec p} \right)\Big]\,.
\end{align}
We give more details of this derivation in appendix~\ref{App:B}.

By defining the vacuum as the state annihilated by the operators,
\begin{equation}
a^{(\lambda)}_{\vec p} |0\rangle = b_{\vec p} |0\rangle =0\,,
\end{equation}
for all $\lambda$, we can define the number operators
associated with the photon and the ghost:
\begin{subequations}
\begin{eqnarray}
\label{eq:number-operator-photon}
N_{\bar A,\lambda }&=& - g_{\lambda\lambda}
a^{ (\lambda)\dag}_{\vec p} a^{ ( \lambda)} _{\vec p}
 \,, \\[2ex]
\label{eq:number-operator-ghost}
N_{G}&=& -b^{\dag}_{\vec p}\, b _{\vec p}\,.
\end{eqnarray}
\end{subequations}
Indeed, the above number operators satisfy the standard relations
\begin{subequations}
\begin{align}
\left[ N_{\bar A,\lambda }, a^{ (\lambda')}_{\vec p}  \right]&= -a^{ (\lambda)}_{\vec p}
 \delta_{\lambda\lambda' }
 \,,    \quad   \left[ N_{\bar A,\lambda }, a^{ ( \lambda')\dag} _{\vec p} \right]=
 a^{ ( \lambda)\dag} _{\vec p} \delta_{\lambda\lambda' } \,, \\[2ex]
\left[ N_{G},b_{\vec p} \, \right]  &= -b_{\vec p}\,,  \qquad \left[ N_{G},
b^{ \dag} _{\vec p} \, \right] =b^{ \dag} _{\vec p} \,.
\end{align}
\end{subequations}
We define $n$-particle states as usual by subsequently
applying creation operators on the vacuum
state:
\begin{subequations}
\begin{align}
|n_{\bar A,\lambda}\rangle&=\frac{1}{\sqrt{n_{\bar A,\lambda}!}}(a^{(\lambda)\dagger}_
{\vec{p}})^{n_{\bar A,\lambda}}|0\rangle\,, \\[2ex]
|n_G\rangle&=\frac{1}{\sqrt{n_G!}}(b^{\dagger}_{\vec{p}})^{n_G}|0\rangle\,,
\end{align}
\end{subequations}
where $n_{\bar A,\lambda}$ is the eigenvalue of the number
operator of Eq.~(\ref{eq:number-operator-photon}) for a
state of $n$ photons of fixed polarization $\lambda$. In an analog
manner, $n_G$ is the eigenvalue of the number
operator of Eq.~(\ref{eq:number-operator-ghost}) for a state of $n$
ghosts. The metric $\eta$
in the state space is given
by the scalar product of such $n$-particle
states~\cite{Lee-Wick:Finite,BG}. For photons, $\langle
n_{\bar A,0}|n_{\bar A,0}\rangle=(-1)^{n_{\bar A,0}}$
for the $\lambda=0$ mode and $\langle
n_{\bar A,k}|n_{\bar A,k}\rangle=1$ for the remaining ones with
$k=1,2$. For ghosts, it holds $\langle n_G|n_G\rangle=(-1)^{n_G}$.
Thus, we see that the states with an odd occupation
number of ghosts have a negative norm.
The metric for the photon can be written as
$\eta_{A,\lambda}=(-g_{\lambda \lambda})^{N_{\bar A,\lambda }}$ with
 $g_{\lambda\lambda'}$ given under Eqs.~(\ref{eq:relations-epsilon})
and that for the ghost reads $\eta_G=(-1)^{N_G}$. Hence, our theory
exhibits an indefinite metric in the
Fock space of the ghost states. It is clear that the same problem occurs
for the $\lambda=0$ mode of the photon, but this
behavior is expected and can be dealt with by the usual Gupta-Bleuler method.

In order to remove the vacuum energy, the normal-ordered Hamiltonian is introduced:
\begin{equation}
\normOrd{H}= \frac12  \int\frac{\mathrm{d}^2\vec{p}}{(2\pi)^2}    \left(
\sum_{\lambda,\lambda'} -g_{\lambda\lambda'} N_{\bar A,\lambda\lambda'}
+ N_{G} \right) \,.
\end{equation}
The latter is positive definite, except for the usual $\lambda=0$ mode of the photon again, which
must be treated with the Gupta-Bleuler formalism. Note that the ghost does lead to issues with the positive
definiteness of the Hamiltonian.
%.................................................................
\subsection{Feynman propagator}
\label{sec:feynman-propagator}
%.................................................................
The next step is to derive the Feynman propagator at the level of field operators
for the theory based on Eq.~(\ref{L-D}) with $\xi=1$.
We employ its definition as the vacuum expectation value of the time-ordered product of field operators
at different spacetime points $x$ and $y$. Hence,
\begin{subequations}
\begin{align}
D^F_{\mu\nu}(x-y)&=\theta(x_0-y_0)  D^{(+)}_{\mu\nu}(x-y) \notag \\
&\phantom{{}={}}+\theta(y_0-x_0)  D^{(-)}_{\mu\nu}(x-y)   \,,
\end{align}
with
\begin{eqnarray}
D^{(+)}_{\mu\nu}(x-y)&=& \langle 0| A_\mu(x)A_\nu(y)|0\rangle\,, \\[2ex]
D^{(-)}_{\mu\nu}(x-y)&=&\langle 0| A_\nu(y)A_\mu(x)|0\rangle \,,
\end{eqnarray}
\end{subequations}
and the Heaviside step function $\theta(x)$.
Using the decomposition of~Eq.~(\ref{planewave}), we define
\begin{eqnarray}
D^F_{\mu\nu}(x-y)= D_{\mu\nu}^{(1)F}(x-y)+ D_{\mu\nu}^{(2)F}(x-y)     \,,
\end{eqnarray}
where the first part,
\begin{subequations}
\begin{align}
D_{\mu\nu}^{(1)F}(x-y)&=\theta(x_0-y_0)  D^{(1)(+)}_{\mu\nu}(x-y) \nonumber \\
&\phantom{{}={}}+\theta(y_0-x_0)  D^{(1)(-)}_{\mu\nu}(x-y)\,,
\end{align}
is the Feynman propagator for photons with
\begin{eqnarray}\label{photon1}
 D_{\mu\nu}^{(1)(+)}(x-y)&=&\langle0| \bar A_\mu (x)\bar A_\nu(y)  |0\rangle  \,,
\\[2ex]
D_{\mu\nu}^{(1)(-)}(x-y)&=&\langle0|  \bar A_\nu(y)  \bar A_\mu (x)    |0\rangle  \,.
\end{eqnarray}
\end{subequations}
Furthermore, the second part is the Feynman propagator of the ghost and it reads
\begin{subequations}
\begin{align}
D_{\mu\nu}^{(2)F}(x-y)&=\theta(x_0-y_0)  D^{(2)(+)}_{\mu\nu}(x-y)
\nonumber \\
&\phantom{{}={}}+\theta(y_0-x_0)  D^{(2)(-)}_{\mu\nu}(x-y)    \,,
\end{align}
where
\begin{eqnarray}\label{ghost2}
 D_{\mu\nu}^{(2)(+)}(x-y)&=&\langle0| G_\mu (x)G_\nu(y)  |0\rangle  \,,
\\[2ex]
D_{\mu\nu}^{(2)(-)}(x-y)&=&\langle0| G_\nu(y)  G_\mu (x)    |0\rangle  \,.
\end{eqnarray}
\end{subequations}
Notice that crossed terms such as $\langle 0|\bar{A}_{\mu}(x)G_{\nu} (y)|0\rangle$ have been set to zero,
since the corresponding field operators commute.

Inserting the field operators of Eqs.~(\ref{photon-field}), (\ref{ghost-field}), we arrive at
\begin{subequations}
\begin{align}
D^{(1)(+)}_{\mu\nu}(z)&=-\int \frac{\mathrm{d}^2\vec{p}}{(2\pi)^2 2\omega_p}
  \sum_{\lambda,\lambda' } g_{\lambda\lambda'} \notag \\
&\phantom{{}={}}\times\bar e_{\mu}^{(\lambda)} ( p) \bar e^{(\lambda')*} _{\nu} ( p) \mathrm{e}^{-\mathrm{i}p\cdot z }\,, \\
D^{(1)(-)}_{\mu\nu}(z)&=-\int \frac{\mathrm{d}^2\vec{p}}{(2\pi)^2 2\omega_p}
   \sum_{\lambda,\lambda' } g_{\lambda\lambda'} \notag \\
&\phantom{{}={}}\times\bar e^{(\lambda)} _{\nu} ( p) \bar e_{\mu}^{(\lambda')*} ( p)  \mathrm{e}^{\mathrm{i}p\cdot z }\,,
\end{align}
\end{subequations}
for the photon and
\begin{subequations}
\begin{eqnarray}
\hspace*{-0.9cm} D^{(2)(+)}_{\mu\nu}(z)&=&-\int \frac{\mathrm{d}^2\vec{p}}{(2\pi)^22\Omega_p}
\bar \varepsilon_{\mu}^{(+)} ( p) \bar \varepsilon^{(+)*}_{\nu} ( p) \mathrm{e}^{-\mathrm{i}p\cdot z }\,,
\\[2ex]
\hspace*{-0.9cm} D^{(2)(-)}_{\mu\nu}(z)&=&-\int \frac{\mathrm{d}^2\vec{p}}{(2\pi)^22\Omega_p}
\bar  \varepsilon^{(+)}_{\nu} ( p)\bar \varepsilon_{\mu}^{(+)*} ( p) \mathrm{e}^{\mathrm{i}p\cdot z }\,,
\end{eqnarray}
\end{subequations}
for the ghost with $z^{\mu}=x^{\mu}-y^{\mu}$. To obtain these results,
we have used the algebra of~Eqs.~(\ref{algebra}), (\ref{algebra-opp}).

In the photon sector, we apply~Eq.~(\ref{tens-T}) to
express the sum over polarization tensors
in terms of the tensor $T_{\mu\nu}$ of Eq.~(\ref{eq:tensor-T}). This leads to
\begin{align}
D_{\mu\nu}^{(1)F}(z)&=- \int \frac{\mathrm{d}^2\vec{p}}{(2\pi)^2 2\omega_p}
\mathrm{e}^{\mathrm{i} \vec p\cdot \vec z}
\left[\theta(z_0)  T_{\mu \nu} (\vec p)      \mathrm{e}^{-\mathrm{i}
\omega_p z_0} \right.  \notag \\
&\phantom{{}={}}\hspace{1.8cm}\left.+\,\theta(-z_0)  T_{\nu \mu} (-\vec p)
 \mathrm{e}^{\mathrm{i} \omega_p z_0}   \right]  \,.
\end{align}
Furthermore, in the ghost sector, we take advantage of relation~(\ref{tens-T-ghost}) to carry out the analogous
steps:
\begin{align}
D_{\mu\nu}^{(2)F}(z)&= \int \frac{\mathrm{d}^2\vec{p}}{(2\pi)^2
2\Omega_p} \mathrm{e}^{\mathrm{i} \vec p\cdot \vec z}
\left[\theta(z_0)  T_{\mu \nu} (\vec p)      \mathrm{e}^{-\mathrm{i}
\Omega_p z_0} \nonumber  \right.  \notag \\
&\phantom{{}={}}\hspace{1.8cm}\left.+\,\theta(-z_0)  T_{\nu \mu} (-\vec p)
 \mathrm{e}^{\mathrm{i} \Omega_p z_0}   \right]  \,.
\end{align}
Now, we consider the following representation of the Heaviside function given by
\begin{eqnarray}\label{theta}
\theta(z_0)=\frac {\mathrm{i}}{2 \pi} \int _{-\infty}^{\infty}\mathrm{d}
\tau\,\frac{\mathrm{e}^{-\mathrm{i}\tau z_0}}{\tau+\mathrm{i}\epsilon}\,,
\end{eqnarray}
where $\epsilon=0^+$ is an infinitesimal, positive parameter.
With the latter representation, we can cast the photon propagator into the
form
\begin{align}
D_{\mu\nu}^{(1)F}(z)&=- \frac {\mathrm{i}}{2 \pi} \int \frac{\mathrm{d}^2\vec{p}}
{(2\pi)^2 2\omega_p} \mathrm{e}^{\mathrm{i} \vec p\cdot \vec z} \notag \\
&\phantom{{}={}} \times\bigg[ \int _{-\infty}^{\infty} \mathrm{d}\tau\, \notag
\frac{\mathrm{e}^{-\mathrm{i}( \omega_p+\tau )z_0}}{\tau+\mathrm{i}\epsilon}  T_{\mu \nu} (\vec p)
\\ & \phantom{{}={}}\hspace{0.5cm} +\int _{-\infty}^{\infty}\mathrm{d}\tau\,\frac{\mathrm{e}^{\mathrm{i}
(\omega_p+\tau) z_0}}{\tau+\mathrm{i}\epsilon}  T_{\nu \mu} (-\vec p)    \bigg]  \,.
\end{align}
Making a change of variables $p_0=\tau+\omega_p $ and  $-p_0
=\tau+\omega_p$ in the first and second integral, respectively, we have
\begin{align}\label{eq:contribution-D1}
D_{\mu\nu}^{(1)F}(z)&=- \mathrm{i} \int \frac{\mathrm{d}^2\vec{p}}{(2\pi)^3 2\omega_p}
\mathrm{e}^{\mathrm{i} \vec p\cdot \vec z} \notag \\
&\phantom{{}={}}\times\int _{-\infty}^{\infty} \mathrm{d}p_0\,\mathrm{e}^{-\mathrm{i}p_0z_0}
 \left[  \frac{T_{\mu \nu}(\vec p,p_0) }{p_0-\omega_p+\mathrm{i}\epsilon}\right. \notag \\
&\phantom{{}={}}\hspace{3cm}\left.- \frac{T_{\nu \mu} (-\vec p,-p_0)    }{p_0+\omega_p-\mathrm{i}\epsilon}     \right] \notag \\
&=- \mathrm{i} \int_{C_F} \frac{\mathrm{d}^3p}{(2\pi)^3} \mathrm{e}^{-\mathrm{i}  p\cdot  z}
\frac{T_{\mu \nu}(p)}{ p^2+\mathrm{i}\epsilon}    \,.
\end{align}
To formulate the final form of the photon propagator, we
benefited from the property $T_{\nu \mu} (-\vec p,-p_0)= T_{\mu \nu} (\vec p,p_0)$.
Furthermore, we have written the integral over $p_0$ as a contour integral
in the complex $p_0$ plane. The contour $C_F$ is closed in the lower half plane
for positive energies and in the upper half plane for negative energies.
It is passed through in counter-clockwise direction.
By evaluating the ghost part in a similar way, we obtain
\begin{align}
\label{eq:contribution-D2}
D_{\mu\nu}^{(2)F}(z)&=\mathrm{i} \int \frac{\mathrm{d}^2\vec{p}}{(2\pi)^3 2\Omega_p}
 \mathrm{e}^{\mathrm{i} \vec p\cdot \vec z} \notag \\
&\phantom{{}={}}\times \int _{-\infty}^{\infty} \mathrm{d}p_0
\,\mathrm{e}^{-\mathrm{i}p_0z_0} \left[  \frac{T_{\mu \nu}(\vec p,p_0) }{p_0-\Omega_p+\mathrm{i}\epsilon}\right. \notag \\
&\phantom{{}={}}\hspace{3cm}\left.- \frac{T_{\nu \mu} (-\vec p,-p_0)    }{p_0+\Omega_p-\mathrm{i}\epsilon}
   \right] \notag \\
&=\mathrm{i} \int_{C_F} \frac{\mathrm{d}^3p}{(2\pi)^2} \mathrm{e}^{-\mathrm{i}
p\cdot  z}\frac{T_{\mu \nu}(p)}{ p^2-g^{-2}+\mathrm{i}\epsilon}    \,,
\end{align}
by writing the integral over $p_0$ as another contour integral
along the same contour $C_F$ introduced before.
Adding the contributions of Eqs.~(\ref{eq:contribution-D1}), (\ref{eq:contribution-D2})
results in
\begin{align}\label{PropX_Pres_ep}
D^{F}_{\mu\nu}(z)&=- \mathrm{i}\int_{C_F}  \frac{\mathrm{d}^3p}{(2\pi)^3}\frac{T_{\mu \nu}(p)}{(p^2+\mathrm{i}\epsilon)
(1-g^2p^2-\mathrm{i}\epsilon)} \notag \\
&\phantom{{}={}}\hspace{2cm}\times  \mathrm{e}^{-\mathrm{i}p\cdot z}\,,
\end{align}
where the infinitesimal parameter $\epsilon$ is only kept at linear order. In momentum space the
Feynman propagator
with the $\mathrm{i}\epsilon$ prescription is
\begin{eqnarray}\label{PropP_Pres_ep}
D^{F}_{\mu\nu}(p)=-  \frac{G_{\mu \nu}(\xi=1,p)}{(p^2+\mathrm{i}\epsilon)
(1-g^2p^2-\mathrm{i}\epsilon)}   \,,
\end{eqnarray}
where we have used~Eq.~(\ref{eq:tensor-T}). The latter can be generalized to arbitrary $\xi$.
By inserting $M=g^{-1}$, we reformulate it as
\begin{equation}
\label{eq:feynman-propagator-general}
D^F_{\mu\nu}(\xi,p)=\frac{M^2G_{\mu\nu}(\xi,p)}{(p^2+\mathrm{i}\epsilon)(p^2-M^2+\mathrm{i}\epsilon)}\,,
\end{equation}
which corresponds to the inverse $P_{\mu\nu}$ of Eq.~(\ref{eq:propagator-pmunu}) for $\epsilon\mapsto 0$.
%.................................................................
\subsection{Microcausality}
%.................................................................
Two spacetime points that cannot be connected by a light signal (or a signal propagating
with lower velocity) are called causally disconnected. In a theory with Lorentz symmetry
intact, such a set of spacetime points is separated by a spacelike interval. When Lorentz
symmetry is violated, the causal structure is not simply determined by the Minkowski metric,
but directly by the propagation velocity of the field operator under consideration, i.e., the
interval need not necessarily be spacelike. As Lorentz symmetry is preserved for our theory,
its causal structure is, indeed, based on the Minkowski metric.

Now, field operators evaluated
at such a set of spacetime points can be considered as independent of each other, i.e., they
should commute. If the latter is the case, microcausality is guaranteed for the theory under
investigation. To prove microcausality for the theory defined by Eq.~(\ref{L-D}),
we start with the basic commutator of field operators at the points $x$ and $y$:
\begin{equation}
D_{\mu \nu }(x-y)=\left[ A_{\mu}(x),  A_{\nu}(y)  \right]\,.
\end{equation}
A direct calculation starting from~Eq.~(\ref{planewave}) provides
\begin{align}
&[{\bar A}_\mu(x),{\bar A}_\nu(y)]=\int \frac{\mathrm{d}^2\vec p\, \mathrm{d}^2\vec k}{(2\pi)^44
\omega_p\omega_{ k}} \notag \\
&\phantom{{}={}}\times\sum_{\lambda,\lambda'}\left(\bar e^{(\lambda)}_\mu(\vec p) \bar e^{*(\lambda') }_\nu(\vec k)
 [a^{(\lambda)}_{\vec p},a^{(\lambda')\dag}_{\vec k}]
\mathrm{e}^{-\mathrm{i}p\cdot x+\mathrm{i}k\cdot y}  \notag   \right.   \\
&\phantom{{}={}}\hspace{0.8cm}\left.+\,\bar e^{*(\lambda)}_\mu(\vec p)\bar e^{(\lambda ')}_\nu(\vec k)
 [a^{(\lambda)\dag}_{\vec p},a^{(\lambda')}_{\vec k}]   \mathrm{e}^{\mathrm{i}p\cdot x-\mathrm{i}k\cdot y}    \right) \,,
\end{align}
and
\begin{align}
&[{G}_\mu(x),{G}_\nu(y)]=\int\frac{\mathrm{d}^2\vec p\,\mathrm{d}^2\vec k}{(2\pi)^4
4 \Omega_p\Omega_k} \notag \\
&\phantom{{}={}}\times\left(\varepsilon^{(+)}_\mu(\vec p)\varepsilon^{(+)*}_\nu (\vec k)[\, b_{\vec p},b^\dag_{\vec k}\,]\,
 \mathrm{e}^{-\mathrm{i}p\cdot x+\mathrm{i}k\cdot y} \right. \nonumber \\
&\phantom{{}={}}\hspace{0.7cm}\left.+\,\varepsilon^{(+)*}_\mu(\vec p)
\varepsilon^{(+)}_\nu(\vec k)[b^\dag_{\vec p},b_{\vec k}]\mathrm{e}^{\mathrm{i}p\cdot x-\mathrm{i}k\cdot y}\right)\,.
\end{align}
Hence, it is important to study the commutator for the photon and the
ghost separately, as the corresponding field operators are independent
of each other. By using the algebra of Eqs.~(\ref{eq:algebra-photon-ghost}) and the properties of the polarization
vectors of~Eqs.~(\ref{tens-T}) and (\ref{tens-T-ghost}), we arrive at
\begin{widetext}
\begin{align}\label{BC}
D_{\mu\nu}(x-y)&=-\int \frac{\mathrm{d}^2\vec p}
{(2\pi)^2}    \frac{ 1 }{2\omega_p } \left( T_{\mu\nu}(\vec p,\omega_p) \mathrm{e}^{-\mathrm{i}p\cdot (x-y)}
 -T_{\nu\mu}(\vec p,\omega_p) \mathrm{e}^{\mathrm{i}p\cdot (x-y)} \right) \notag \\
&\phantom{{}={}}+\int \frac{\mathrm{d}^2\vec p}
{(2\pi)^2}    \frac{ 1}{2 \Omega_p } \left( T_{\mu\nu} (\vec p,\Omega_p)
\mathrm{e}^{-\mathrm{i}p\cdot (x-y)} -T_{\nu\mu} (\vec p,\Omega_p) \mathrm{e}^{\mathrm{i}p\cdot (x-y)}  \right) \,,
\end{align}
\end{widetext}
where we employed the tensor $T_{\mu\nu}$ of~Eq.~(\ref{eq:tensor-T}).
We define $z=x-y$ and perform a change of variables $\vec p \to -\vec p$
in the second contribution above, to obtain
\begin{widetext}
\begin{align}
D_{\mu \nu }(z)&=-\int \frac{\mathrm{d}^2\vec p}
{(2\pi)^2}    \frac{ \mathrm{e}^{\mathrm{i}\vec p\cdot \vec z} }{2\omega_p }
\left( T_{\mu\nu}(\vec p,\omega_p) \mathrm{e}^{-\mathrm{i}\omega_p z_0}
 -T_{\nu\mu}(-\vec p,\omega_p) \mathrm{e}^{\mathrm{i}\omega_p z_0}\right)
\notag \\
&\phantom{{}={}}+\int \frac{\mathrm{d}^2\vec p}
{(2\pi)^2}    \frac{ \mathrm{e}^{\mathrm{i}\vec p\cdot \vec z}}{2 \Omega_p } \left( T_{\mu\nu} (\vec p,\Omega_p)
\mathrm{e}^{-\mathrm{i} \Omega_p z_0} -T_{\nu\mu} (-\vec p,\Omega_p) \mathrm{e}^{\mathrm{i}\Omega_p z_0} \right) \,.
\end{align}
\end{widetext}

Since $T_{\nu\mu}(-\vec p,p_0)=T_{\mu\nu}(\vec p,-p_0)$, we can introduce
another contour integral in the complex $p_0$ plane along a contour $C$ that encircles all poles in
counter-clockwise direction:
\begin{align}
D_{\mu \nu }(z)&=\mathrm{i} \int \frac{\mathrm{d}^2\vec p}
{(2\pi)^2}\int_{C}  \frac{\mathrm{d}p_0}
{2\pi}   \left( \frac{T_{\mu\nu}   (\vec p,p_0)}{(p_0+\omega_p)  (p_0-\omega_p)  }
 \right. \nonumber \\
&\phantom{{}={}}-\left.\frac{T_{\mu\nu} (\vec p,p_0)}{(p_0+\Omega_p)
  (p_0-\Omega_p)  }\right) \mathrm{e}^{-\mathrm{i}p\cdot z}\,.
\end{align}
Note that the contour $C$ is different from the contour $C_F$ that we defined in the context
of the Feynman propagator in Sec.~\ref{sec:feynman-propagator}. Therefore,
\begin{equation}
D_{\mu \nu }(z)=\mathrm{i}\int _{C} \frac{\mathrm{d}^3 p}
{(2\pi)^3}    \frac{T_{\mu\nu}   (p)}{p^2 (1-g^2p^2)  }  \mathrm{e}^{-\mathrm{i}p\cdot z}\,.
\end{equation}

To prove that this expression vanishes outside the light cone, that is, for $(x-y)^2<0$,
we can perform a Lorentz transformation of the coordinates to a frame where $x^0-y^0=0$ and
compute the
integral in this new frame. Thus, we focus on the integral over $p_0$,
\begin{align}
I_{\mu \nu}&=\int_C \mathrm{d}p_0 \,\frac{T_{\mu\nu}(p) }{p^2(1-g^2p^2)}
\notag \\   \nonumber
&=-\frac{1}{g^2} \int_C \mathrm{d}p_0\,T_{\mu\nu}(p)\left[(p_0-\omega_p)(p_0+\omega_p)\right. \notag \\
&\phantom{{}={}}\hspace{1.8cm}\left.\times(p_0-\Omega_p)(p_0+\Omega_p)\right]^{-1}\,,
\end{align}
whose result is given by
\begin{align}
-g^2I_{\mu \nu}&=
\left[\frac{T_{\mu\nu}(\vec p,\omega_p) }{2 \omega_p  ( \omega_p^2-  \Omega_p^2)  }-
\frac{T_{\mu\nu}(\vec p,-\omega_p)}{  2 \omega_p  ( \omega_p^2-  \Omega_p^2)    }\right] \notag \\
&\phantom{{}={}}+\left[\frac{T_{\mu\nu}(\vec p,\Omega_p)}{  2 \Omega_p  ( \Omega_p^2-  \omega_p^2)   }
-\frac{T_{\mu\nu}(\vec p,-\Omega_p)}{  2 \Omega_p  ( \Omega_p^2-  \omega_p^2)    }\right] \,.
\end{align}%%
Now we employ the explicit form of the tensor $T_{\mu\nu}$ in~Eq.~(\ref{eq:tensor-T}).
The terms proportional to $\eta_{\mu \nu}$ cancel for each contribution enclosed in parentheses
as well as those proportional to $g^2p_ip_j $ and $\mathrm{i}g \epsilon_{0ij}p^i$. The only terms that survive
are proportional to $g^2p_0p_i$ and $\mathrm{i}g \epsilon_{i0j}p_0$. However, these cancel due to the
identity
\begin{equation}
\frac{\omega_p}{\omega_p  ( \omega_p^2-  \Omega_p^2)}+\frac{\Omega_p}{\Omega_p  ( \Omega_p^2-  \omega_p^2)}=0\,,
\end{equation}%%
whereupon $I_{\mu\nu}=0$ and $D_{\mu\nu}=0$ in the particular frame considered. Lorentz
invariance allows us to generalize this finding to an arbitrary frame.
We conclude that the theory is microcausal, since the commutator of two field operators vanishes
when they are evaluated at causally disconnected spacetime points.
%%%%%%%%%%%%%%%%%%%%%%%%%%%%%%
\section{Perturbative unitarity} \label{sec4}
%%%%%%%%%%%%%%%%%%%%%%%%%%%%%%
In the previous sections we have seen that the theory defined by (\ref{L-D}) develops an
indefinite metric in the Hilbert space of states
due to higher-time derivatives present in the Lagrangian. This metric
 is responsible for
negative-norm states and could possibly induce a violation of unitarity.
As a consequence of this, the normal probabilistic interpretation of quantum theory
would be undermined.

Unitarity can be investigated in various ways. A reasonable method
for a free theory is to study the condition
of reflection positivity~\cite{Montvay:1994}. However, in the presence
of interactions,
computations based on the optical theorem in perturbation theory~\cite{Peskin:1995} are better
under control. In this context, imaginary parts of forward scattering
amplitudes are compared to cross sections
of processes corresponding to cut Feynman diagrams. In the forthcoming
subsections we check the validity of unitarity of the theory via reflection positivity and the optical theorem.
%----------------------------------------------------------------
\subsection{Reflection positivity}
%----------------------------------------------------------------
Reflection positivity is a property of a scalar two-point function in Euclidean
space that guarantees the validity of unitarity of the corresponding free field theory in
Minkowski spacetime. It is primarily used in the context of lattice gauge theory, but also
found application in proofs of unitarity for Lorentz-violating theories (see,
e.g.,~\cite{Adam:2001ma,Klinkhamer:2010zs,Schreck:2014qka} for applications to
Maxwell-Chern-Simons theory in $(3+1)$ dimensions, modified Maxwell theory, and
higher-derivative theories of fermions).

\begin{figure}[t]
\centering
\includegraphics[width=0.42\textwidth]{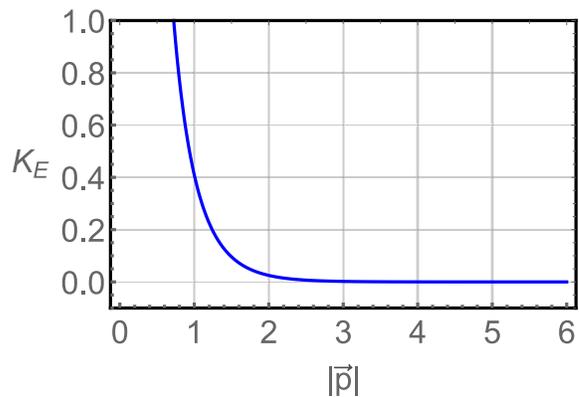}
\caption{\label{Fig1} Plot of the function $K_E(x_3,|\vec p|)$ of
Eq.~(\ref{eq:fourier-transformed-wick-rotated-propagator}) for $x_3=2$ and $M=2$
as a function of $|\vec{p}|$.}
\end{figure}

To check the validity of reflection positivity for our theory, we
will make some simplifications as follows.
Let us consider the combination of poles in the scalar propagator function
\begin{equation}
K(p_0,\vec p)=\frac{M^2}{p^2(p^2-M^2)}\,,
\end{equation}
whose form is taken from~Eq.~(\ref{prop_xi}). We can rearrange the latter as
\begin{eqnarray}
K(p_0,\vec p)&=&-\frac{1}{p^2}+\frac{1}{p^2-M^2}\,.
\end{eqnarray}
Now we go to Euclidean space by means of the replacement $p_0\rightarrow \mathrm{i}p_3$
\begin{equation}
K(p_0,\vec{p})\mapsto K_E(p_3,\vec p)=\frac{1}{p_E^2}-\frac{1}{p_E^2+M^2}\,.
\end{equation}
The weak version of reflection positivity requires that the one-dimensional Fourier
transform of the latter Euclidean propagator function with respect to $p_3$ be
nonnegative. Computing this Fourier transform leads to
\begin{align}
\label{eq:fourier-transformed-wick-rotated-propagator}
K_E(x_3,|\vec p|)&=\int_{-\infty}^{\infty}\mathrm{d}p_3\,K_E(p_3,|\vec{p}|)
\mathrm{e}^{-\mathrm{i}p_3x_3} \notag \\
&=\int_{-\infty}^\infty \mathrm{d}p_4\,\frac{\mathrm{e}^{-\mathrm{i}
p_3x_3}}{p_3^2+\vec{p}^{\,2}}-\int_{-\infty}^\infty \mathrm{d}p_3
\,\frac{\mathrm{e}^{-\mathrm{i}p_3x_3}}{p_3^2+\vec{p}^{\,2}+M^2} \notag \\
&=\pi\left[\frac{\exp(-|x_3||\vec p|)}{|\vec p|}\right. \notag \\
&\phantom{{}={}}\hspace{0.4cm}\left.-\frac{\exp\left(-|x_3|\sqrt{\vec p^{\,2}+M^2}\right)}{\sqrt{\vec p^{\,2}+M^2}}\right]\,.
\end{align}
We see that the latter expression is nonnegative for all momentum magnitudes
$|\vec{p}|$ (see~Fig.~\ref{Fig1}). However, it should be noted that the condition of reflection
positivity refers to the scalar part of the two-point function only. Also, it does not
take into account interactions. Therefore, reflection positivity does not provide a complete
understanding of unitarity when the tensor structure of the two-point function and interactions
are taken into consideration.

To check the validity of unitarity more thoroughly, it is wise to go
beyond this technique and, for instance, use the optical theorem. In the next section, we give
an example in which a study of the optical theorem with the complete structure of poles and
polarization vectors is indispensable.
%...................................................Fig1......................
%%
%................................................................................
%----------------------------------------------------------------
\subsection{Electron-positron annihilation at tree-level}
%----------------------------------------------------------------
Our intention is to check the perturbative validity of the optical theorem
for the theory defined by Eq.~(\ref{L-D}). To do so, we have to couple the modified photon theory
to standard Dirac fermions in $(2+1)$ dimensions, i.e., we will consider a modified QED in three
dimensions (QED${}_3$). A summary on a theory of Dirac spinors in $(2+1)$ dimensions is given
in appendix~\ref{App:D}. We then write the total Lagrange density as
\begin{subequations}
\begin{align}
\mathcal{L}_{\mathrm{tot}}&=\mathcal{L}+\mathcal{L}_{\psi,\upgamma}\,, \displaybreak[0]\\[2ex]
\mathcal{L}_{\psi,\upgamma}&=\overline{\psi}[\gamma^{\mu}(\mathrm{i}\partial_{\mu}-eA_{\mu})+m]\psi\,,
\end{align}
\end{subequations}
with $\mathcal{L}$ given by Eq.~(\ref{L-D}). Here, $e$ is the electric charge,
$m$ the fermion mass, $\psi$ is a four-component Dirac spinor,
and $\gamma^{\mu}$ the set of three Dirac matrices of Eq.~(\ref{eq:dirac-matrices}). Note again that
Lorentz indices run over $0,1,2$.

\begin{figure}[b]
\centering
\includegraphics[width=0.4\textwidth]{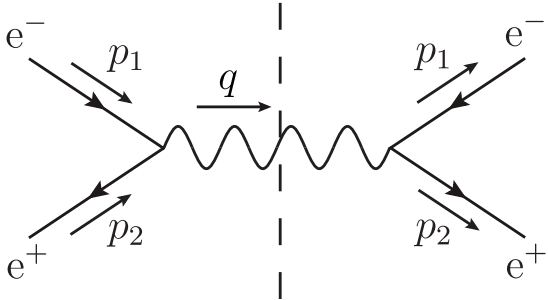}
\caption{\label{Fig2} Polarized forward scattering electron-positron annihilation where a cut of the
gauge field propagator is indicated by the dashed line. The three-momenta of the incoming particles
are $p_1$, $p_2$ where the three-momentum of the intermediate modified photon is denoted as $q$.}
\end{figure}

The optical theorem establishes a connection between
the forward-scattering amplitude of a particular particle physics process and the decay rates
or total cross sections of processes that are obtained by cutting the Feynman diagram of the
forward-scattering amplitude into two pieces. We will study processes at tree-level and
one-loop order that involve the gauge-field propagator (\ref{prop_xi}) of
the theory~\cite{Bhabha1,Bhabha2}. Let us start with the polarized forward scattering annihilation process of electron-positron pairs,
$\mathrm{e^+e^-}\to \mathrm{e^+e^-}$ of Fig.~\ref{Fig2}.
%...................................................Fig2......................
%%
%................................................................................
The corresponding amplitude is given by
\begin{align}
\mathrm{i}\mathcal{M}_F&=\bar{v}(p_2)(-\mathrm{i}e\gamma^\mu)u(p_1)\left(\mathrm{i}
D^F_{\mu\nu}(\xi,q) \right) \notag \\
&\phantom{{}={}}\times\bar{u}(p_1)(-\mathrm{i}e\gamma^\nu)v(p_2)\,,
\end{align}
with the Feynman propagator of Eq.~(\ref{eq:feynman-propagator-general}) and $q=p_1+p_2$. Particle and antiparticle
spinors of a particular spin projection are denoted as $u(p)$ and $v(p)$, respectively, and
correspond to those of Eqs.~(\ref{eq:particle-spinors}), (\ref{eq:antiparticle-spinors}). Considering
polarized scattering is not crucial for the verification
of the optical theorem, though. It just simplifies the expressions, as the polarizations of the incoming and
outgoing particles need not be averaged or summed over. Note also that we suppress the spin index for
external spinors. Now, we can write
\begin{equation}
\label{eq:forward-scattering-amplitude-electron-positron}
\mathcal{M}_F= -e^2 \mathcal{M}^\mu(p_1,p_2)
D^F_{\mu\nu}(\xi,q) \mathcal{M}^{\dag \nu}(p_1,p_2)  \,,
\end{equation}
where
\begin{eqnarray}
\mathcal{M}^\mu(p_1,p_2)&=&\bar{v}(p_2)\gamma^\mu u(p_1)  \,, \\[1ex]
\mathcal{M}^{\dag \nu}(p_1,p_2)&=& \bar{u}(p_1) \gamma^\nu v(p_2)  \,.
\end{eqnarray}
The process that results from cutting the diagram of the forward scattering amplitude into two
pieces is the production of a modified photon by an electron positron pair. In contrast to what
happens in standard QED, the cross section of this process is not necessarily equal to zero due
to energy-momentum conservation. The reason is the presence of the massive ghost, which can render
the process possible. In this case, the condition of energy conservation can be evaluated
in the center-of-mass frame: $|\vec p_1|= |\vec p_2|=1/2g$. Therefore, it will be sufficient to
prove unitarity by considering the contributions to the imaginary part (or discontinuity) of the
amplitude for the massive ghost.

In the forward scattering amplitude of~Eq.~(\ref{eq:forward-scattering-amplitude-electron-positron})
an integral over the three-momentum $q$ of the intermediate state can be introduced that is canceled
again by the three-dimensional $\delta$ function of total energy-momentum conservation (which is
equivalent to energy-momentum conservation at each vertex):
\begin{align}
\mathcal{M}_F&=-e^2 \int \frac{\mathrm{d}^3q }{(2\pi)^3 }  \mathcal{M}^\mu
D^F_{\mu\nu}(\xi,q) \mathcal{M}^{\dag \nu} \notag \\
&\phantom{{}={}}\times (2\pi)^3\delta^{(3)} (p_1+p_2-q)  \,.
\end{align}
By inserting the Feynman propagator of Eq.~(\ref{eq:feynman-propagator-general}), we have
\begin{align}
\mathcal{M}_F&=-e^2 M^2 \int \frac{\mathrm{d}^3q}{(2\pi)^3} \frac{ \mathcal{M}^\mu
G_{\mu\nu} (\xi,q) \mathcal{M}^{\dag \nu}}{(q^2+\mathrm{i}\epsilon)(q^2-M^2+\mathrm{i}\epsilon)}
\nonumber \\
&\phantom{{}={}}\times(2\pi)^3 \delta^{(3)}(p_1+p_2-q)\,.
\end{align}
As the photon propagator is coupled to a conserved external current and energy-momentum is conserved at the vertex, we can use the Ward identity to get rid
of all terms in the propagator proportional to this momentum: $q_{\mu} \mathcal{M}^{\mu}=0$.
Doing so, allows for instating the tensor $T_{\mu \nu}$ of Eq.~(\ref{eq:tensor-T}).

It is valuable to recall that
\begin{equation}\label{M-rel}
\frac{  M^2 }{(q^2+\mathrm{i}\epsilon )(q^2-M^2+\mathrm{i}\epsilon)}
=-\frac{  1 }{q^2+\mathrm{i}\epsilon}  +
\frac{  1 }{q^2-M^2+\mathrm{i}\epsilon} \,.
\end{equation}
By making use of the latter, we can decompose the denominator into two parts:
\begin{align}
\mathcal{M}_F&=e^2\int\frac{\mathrm{d}^3q }{(2\pi)^3 } \left[  \frac{  \mathcal{M}^\mu
T_{\mu\nu} (q) \mathcal{M}^{\dag \nu}  }{q^2+\mathrm{i}\epsilon}-\frac{  \mathcal{M}^\mu
T_{\mu\nu} (q) \mathcal{M}^{\dag \nu}  }{q^2-M^2+\mathrm{i}\epsilon}   \right] \nonumber \\
&\phantom{{}={}}\times (2\pi)^3  \delta^{(3)} (p_1+p_2-q)  \,.
\end{align}
Now we insert the expression for $T_{\mu \nu}$ in terms of polarization vectors given
in~Eqs.~(\ref{tens-T}), (\ref{tens-T-ghost}) and obtain
\begin{align}
\mathcal{M}_F&=e^2  \int \frac{\mathrm{d}^3q }{(2\pi)^3 } \left[  \frac{
\sum_{\lambda,\lambda'}  \left( \mathcal{M}^{\mu} \bar e^{(\lambda)}_{\mu}
\right) g_{ \lambda\lambda'}
\left(\mathcal{M}^{\dag \nu}    \bar e^{(\lambda')*}_{\nu}\right)  }{q^2+\mathrm{i}\epsilon}
\nonumber  \right. \\
&\phantom{{}={}}\hspace{2cm}\left.+\frac{\left(\mathcal{M}^{\mu} \bar  \varepsilon^{(+)}_{\mu} \right)
\left(\mathcal{M}^{\dag \nu} \bar  \varepsilon^{(+)*}_{\nu} \right)}{q^2-M^2+\mathrm{i}\epsilon}\right] \nonumber \\
&\phantom{{}={}}\hspace{2cm}\times (2\pi)^3  \delta^{(3)} (p_1+p_2-q)  \,.
\end{align}
Since it is not possible to satisfy energy-momentum conservation and the
dispersion relation for the photon at the same time, the first contribution is zero. We are then left with
\begin{align}
\mathcal{M}_F&=e^2  \int \frac{\mathrm{d}^3q }{(2\pi)^3 } \frac{    | \mathcal{M}^{\mu}
 \bar \varepsilon^{(+)}_{\mu} (q)|^2  }{(q_0+\Omega_q-\mathrm{i}\epsilon)
  (q_0-\Omega_q+\mathrm{i}\epsilon)} \nonumber  \\
&\phantom{{}={}}\times   (2\pi)^3 \delta^{(3)} (p_1+p_2-q)  \,.
\end{align}
We perform the integration over $q_0$ by defining the center-of mass
energy $\sqrt{s}=p_1^0+p_2^0$ and exploit the property of the $\delta$ function.
This leads to
 \begin{align}
 \mathcal{M}_F(s)&=e^2  \int \frac{\mathrm{d}^2\vec q }{(2\pi)^3 } \frac{
   |\mathcal{M}^{\mu} \bar \varepsilon^{(+)}_{\mu}
    (\Omega_q,\vec q)|^2  }{(\sqrt{s}+\Omega_q-\mathrm{i}\epsilon)
    (\sqrt{s}-\Omega_q+\mathrm{i}\epsilon)}  \nonumber \\
  &\phantom{{}={}}\times  (2\pi)^3
     \delta^{(2)} (\vec p_1+\vec p_2-\vec q)  \,.
\end{align}
The imaginary part of the amplitude can be evaluated based on
the identity
\begin{eqnarray}\label{im-part}
\lim_{\epsilon\rightarrow0^+} \frac{1}{x\pm \mathrm{i}\epsilon}
=\mathcal{P}\left(\frac1x \right)  \mp \mathrm{i}\pi \delta(x)\,,
\end{eqnarray}
where $\mathcal{P}$ denotes the principal value. We also consider
 \begin{align}
& \frac{  2 \Omega_q  }{(\sqrt{s}+\Omega_q-\mathrm{i}\epsilon)
  (\sqrt{s}-\Omega_q+\mathrm{i}\epsilon)}  \nonumber  \\ & = \frac{  1  }{\sqrt{s}-\Omega_q
  +\mathrm{i}\epsilon } - \frac{  1  }{ \sqrt{s}+\Omega_q-\mathrm{i}\epsilon }   \,.
\end{align}
The result is
\begin{align}
\label{eq:imaginary-part-amplitude-tree-level}
\text{Im}( \mathcal{M}_F(s))&=-e^2 \int \frac{\mathrm{d}^2\vec q}{(2\pi)^3}
|\mathcal{M}^{\mu}\bar \varepsilon^{(+)}_{\mu} (\Omega_q,\vec q)|^2 \notag \\
&\phantom{{}={}}\times (2\pi)^3 \delta^{(2)} (\vec p_1+\vec p_2-\vec q) \notag \\
&\phantom{{}={}}\times \frac{\pi}{2\Omega_q}\left[\delta(\sqrt{s}-\Omega_p)  +
 \delta(\sqrt{s}+\Omega_q)     \right]\,.
\end{align}
The second $\delta$ function in Eq.~(\ref{eq:imaginary-part-amplitude-tree-level})
involves a non-zero contribution coming from the
possibility of negative energies. This can be seen in the following way.
From the definition of the Feynman propagator one has
\begin{align}
D^F_{\mu\nu}(z_0,\vec z)&=\theta(z_0)D^{(+)}_{\mu\nu}(z_0,\vec z) \notag \\
&\phantom{{}={}}+\theta(-z_0)D^{(-)}_{\mu\nu}(z_0,\vec z)\,,
\end{align}
Performing a coordinate Poincar\'{e} transformation, for instance,
a constant time translation that adds a constant
purely timelike three-vector
to $z$ such that $z_0\to -z_0$, one has
\begin{align}
D^F_{\mu\nu}(-z_0,\vec z)&=\theta(-z_0)D^{(+)}_{\mu\nu}(-z_0,\vec z) \notag \\
&\phantom{{}={}}+\theta(z_0)  D^{(-)}_{\mu\nu}(-z_0,\vec z) \,.
\end{align}
The interpretation is that negative energies occur in the opposite
flow of time. This is precisely the reason why we include the
second $\delta$ function in Eq.~(\ref{eq:imaginary-part-amplitude-tree-level}).
In the literature, the latter is sometimes
represented by a cut with a shaded region indicating the corresponding
 direction of energy flow~\cite{Perturbativeunitarity}.
%...................................................Fig3......................
%\begin{figure}[b]
\begin{figure}[t]
\centering
\includegraphics[width=0.42\textwidth]{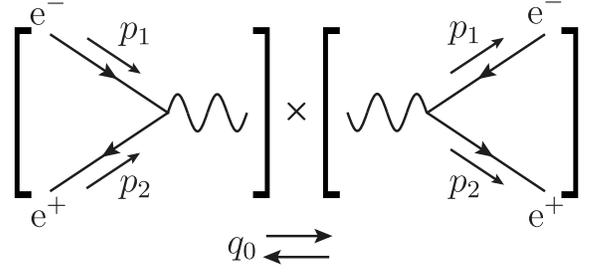}
\caption{\label{Fig3} After cutting the photon propagator in the diagram of Fig.~\ref{Fig2}, the sum over intermediate
states in both directions of the energy flow is considered.}
\end{figure}
%................................................................................
Finally, we can write
\begin{align}
\label{final_result}
2\text{Im}(\mathcal{M}_F(s))&=-e^2\int\frac{\mathrm{d}^3q }{(2\pi)^3}|\mathcal{M}^{\mu}\bar\varepsilon^{(+)}_{\mu}(q)|^2 \notag \\
&\phantom{{}={}}\times (2\pi)^3 \delta^{(3)}(p_1+p_2-q)(2\pi)\delta(q^2-M^2) \notag \\
&\phantom{{}={}}\times \left[\theta(q_0)+\theta(-q_0)\right]\,,
\end{align}
which represents the sum of diagrams with energy flow in the positive
and negative direction as represented in~Fig.~\ref{Fig3}.

Now we come to the crucial point in the analysis where we must introduce some of the ideas
developed by Lee and Wick. As a first observation,
the negative global sign in Eq.~(\ref{final_result}) may threaten unitarity
since the left-hand side of the latter equation, which is related to the cross section, is positive-definite.
To overcome this problem, we apply the Lee-Wick prescription that removes the negative-metric states from the asymptotic
Hilbert space. Furthermore, as long as the energy of the incoming state is low enough, the argument of
$\delta(q^2-M^2)$ is impossibly different from zero due to the large mass of the ghost. The latter will simply not be excited
under this condition. Then unitarity is guaranteed in a direct way just as in the
standard case~\cite{Lee-Wick:Negative,Lee-Wick:Finite,Anon-analytic}.\\

%----------------------------------------------------------------
\subsection{Compton scattering at one-loop level}
%----------------------------------------------------------------
Our next step is to study unitarity when virtual ghosts arise in loop diagrams.
We analyze the optical theorem for the (polarized) Compton scattering process of Fig.~\ref{Fig4}.
The forward-scattering amplitude at one-loop level for this process in the
extended Maxwell-Chern-Simons theory in $(2+1)$
dimensions given by the Lagrangian (\ref{L-D})
reads
\begin{widetext}
\begin{align}
\label{eq:one-loop-amplitude-compton}
\mathrm{i}\mathcal{M}&=\epsilon^{ (\lambda) \ast}_\beta(k  )\bar{u}(p')
(-\mathrm{i}e\gamma^\beta) \left(\frac{\mathrm{i}(\slashed{p}+m)}
{p^2-m^2+\mathrm{i}\epsilon}\right) (-\mathrm{i}e\gamma^\mu)  \int \frac{\mathrm{d}^3q}{(2\pi)^3}
\frac{\mathrm{i}(\slashed{p}-\slashed{q}+m)}{(p-q)^2-m^2+\mathrm{i}\epsilon}  \nonumber  \\
&\phantom{{}={}}\times (\mathrm{i}D^F_{\mu\nu}(q)) (-\mathrm{i}e\gamma^\nu)  \left(\frac{\mathrm{i}(\slashed{p}
+m)}{p^2-m^2+\mathrm{i}\epsilon}\right)
  (-\mathrm{i}e\gamma^\alpha) u(p')\epsilon^{(\lambda)}_\alpha(k) \,,
\end{align}
where the external electrons and photons are considered as polarized.
The fermion propagator for $(2+1)$-dimensional Dirac theory of
Eq.~(\ref{eq:fermion-propagator}) has been inserted.
For simplicity, we choose the particular gauge fixing parameter $\xi=1$
and employ the Feynman propagator of Eq.~(\ref{PropP_Pres_ep}).
We introduce the following short-hand notation for expressions
formed from external spinors and polarization vectors:
\begin{subequations}
\begin{align}
J_1^{(\lambda)} (p',k)&=\epsilon^{ \ast(\lambda)}_\beta(k  )\bar{u}(p')
 \gamma^\beta \,, \\[2ex]
J_2^{(\lambda)}  (p',k)&= \gamma^\alpha u(p')\epsilon^{(\lambda)}_\alpha(k)  \,,
\end{align}
\end{subequations}
and rewrite the denominators of~Eq.~(\ref{eq:one-loop-amplitude-compton})
in terms of the poles. We also work in the center-of-mass frame where $\vec p=\vec{0}$ and
use~Eq.~(\ref{M-rel}) to obtain
\begin{align}
\label{eq:forward-scattering-compton}
\mathrm{i} \mathcal{M}&=-e^4 J_1^{(\lambda)}  (p',k) \left(\frac{\slashed{p}+m}{p^2-m^2+\mathrm{i}\epsilon}
\right)  \gamma^\mu
  \int \frac{\mathrm{d}^3q}{(2\pi)^3}
\frac{\slashed{p}-\slashed{q}+m}{(q_0-p_0-E_{q}+\mathrm{i}\epsilon) (q_0-p_0+E_{q}-\mathrm{i}\epsilon)  }
  \notag \\
&\phantom{{}={}}\times  T_{\mu \nu} (q) \left[  \frac{    1}{  (q_0-\omega_q
+\mathrm{i}\epsilon)  (q_0+\omega_q-\mathrm{i}\epsilon)  }
-\frac{  1 }{  (q_0-\Omega_q+\mathrm{i}\epsilon)
 (q_0+\Omega_q-\mathrm{i}\epsilon)   }   \right] \notag \\
&\phantom{{}={}}\times \gamma^\nu \left(\frac{\slashed{p}+m}{p^2-m^2
+\mathrm{i}\epsilon}
\right) J_2^{(\lambda)} (p',k)\,.
\end{align}
Let us decompose the amplitude into a sum of amplitudes via
\begin{subequations}
\begin{eqnarray}
\mathcal{M}&=& \mathcal{M}^{(1)}+\mathcal{M}^{(2)}\,,
\end{eqnarray}
with
\begin{align}
\mathrm{i} \mathcal{M}^{(1)}&=-e^4 J_1^{(\lambda)} (p',k)   \left(\frac{\slashed{p}+m}{p^2-m^2+\mathrm{i}\epsilon}
\right)    \gamma^\mu
  \int \frac{\mathrm{d}^2\vec q \, \mathrm{d}q_0}{(2\pi)^3}
\frac{\slashed{p}-\slashed{q}+m}{(q_0-p_0-E_{q}+\mathrm{i}\epsilon) (q_0-p_0+E_{q}-\mathrm{i}\epsilon)  }
\notag \\
&\phantom{{}={}}  \times \frac{  T_{\mu \nu} (q)  }{  (q_0-\omega_q+\mathrm{i}\epsilon)  (q_0+\omega_q-\mathrm{i}\epsilon)  }
    \gamma^\nu\left(\frac{\slashed{p}+m}{p^2-m^2+\mathrm{i}\epsilon}
\right)  J_2^{(\lambda)}  (p',k)  \,,
\end{align}
and
\begin{align}
\mathrm{i} \mathcal{M}^{(2)}&=e^4 J_1 ^{(\lambda)} (p',k)
 \left(\frac{\slashed{p}+m}{p^2-m^2+\mathrm{i}\epsilon}
\right)    \gamma^\mu
  \int \frac{\mathrm{d}^2\vec q \, \mathrm{d}q_0}{(2\pi)^3}
\frac{\slashed{p}-\slashed{q}+m}{(q_0-p_0-E_{q}+\mathrm{i}\epsilon)
 (q_0-p_0+E_{q}-\mathrm{i}\epsilon)  }
\nonumber   \\
&\phantom{{}={}}  \times
\frac{   T_{\mu \nu} (q) }{  (q_0-\Omega_q+\mathrm{i}\epsilon)
  (q_0+\Omega_q-\mathrm{i}\epsilon)   }
    \gamma^\nu \left(\frac{\slashed{p}+m}{p^2-m^2+\mathrm{i}\epsilon}
\right)  J_2^{(\lambda)}  (p',k)   \,.
\end{align}
\end{subequations}
Our next step is to integrate over the complex variable $q_0$ by using the residue theorem
and closing the contour in the lower half plane of the complex $q_0$
plane. Each integrand has two contributing poles leading to four poles $q_0=z_i$ ($i=1\dots 4$),
in total. For the first integrand we have
\begin{subequations}
\begin{align}
z_1&=p_0+E_q-\mathrm{i}\epsilon\,, \\[1ex]
z_2&=\omega_q-\mathrm{i}\epsilon\,,
\end{align}
\end{subequations}
%...................................................Fig4......................
\begin{figure}[t]
\centering
\includegraphics[width=0.42\textwidth]{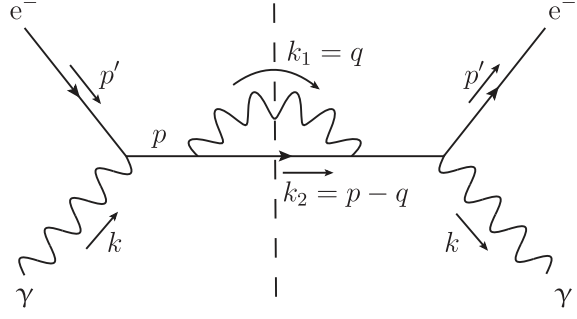}
\caption{\label{Fig4} Forward Compton scattering with one-loop
correction of the fermion propagator included. The cut of both
propagators is indicated by a dashed line. The external three-momenta are given by $k$ and $p'$.}
\end{figure}
%----------------------------------------------------------------
where $E_q$ is the dispersion relation (\ref{eq:fermion-dispersion-relation}) of a
massive fermion in $(2+1)$ dimensions. The poles
of the second integrand are given by
\end{widetext}
\begin{subequations}
\begin{align}
z_3&=p_0+E_q-\mathrm{i}\epsilon\,, \\[1ex]
z_4&=\Omega_q-\mathrm{i}\epsilon\,.
\end{align}
\end{subequations}
We then arrive at

\begin{align}
\mathcal{M}^{(1)}&=-e^4 J_1 ^{(\lambda)}  (p',k)   \left(\frac{\slashed{p}+m}{p^2-m^2}
\right)\gamma^\mu
\nonumber \\
&\phantom{{}={}}\times \int\frac{\mathrm{d}^2\vec q }{(2\pi)^2}
(\slashed{p}-\slashed{q}+m) T_{\mu \nu} (q)
\nonumber \\
&\phantom{{}={}}\times \left(\mathrm{Res}(z_1)+\mathrm{Res}(z_2)\right)
\gamma^\nu\left(\frac{\slashed{p}+m}{p^2-m^2}
\right) \notag \\
&\phantom{{}={}}\times J_2^{(\lambda)}(p',k)\,,
\end{align}
and
\begin{align}
\mathcal{M}^{(2)}&=e^4 J_1 ^{(\lambda)}(p',k)\left(\frac{\slashed{p}+m}{p^2-m^2}\right)\gamma^\mu \nonumber \\
&\phantom{{}={}} \times \int\frac{\mathrm{d}^2\vec q }{(2\pi)^2}
(\slashed{p}-\slashed{q}+m)T_{\mu \nu}(q)
\nonumber  \\
&\phantom{{}={}}\times \left(\mathrm{Res}(z_3)+\mathrm{Res}(z_4)\right)
\gamma^\nu\left(\frac{\slashed{p}+m}{p^2-m^2}
\right) \notag \\
&\phantom{{}={}}\times J_2 ^{(\lambda)}(p',k)\,,
\end{align}
with the residues
\begin{subequations}
\begin{align}
\mathrm{Res}(z_1)&=\frac{-1 }{2E_q (p_0+E_q-\omega_q)
(p_0+E_q+\omega_q-\mathrm{i}\epsilon)  }\,, \displaybreak[0]\\[2ex]
\mathrm{Res}(z_2)&=\frac{-1  }{2\omega_q (p_0+E_q-\omega_q)
 (p_0-E_q-\omega_q+\mathrm{i}\epsilon)  }\,,
\end{align}
\end{subequations}
and
\begin{subequations}
\begin{align}
\mathrm{Res}(z_3)&= \frac{-1  }{2E_q (p_0+E_q-\Omega_q)
 (p_0+E_q+\Omega_q-\mathrm{i}\epsilon)  }\,, \displaybreak[0]\\[2ex]
\mathrm{Res}(z_4)&= \frac{-1  }{2\Omega_q
(p_0+E_q-\omega_q)  (p_0-E_q-\Omega_q+\mathrm{i}\epsilon)  }\,,
\end{align}
\end{subequations}
where we have rescaled the parameter $\epsilon$ and set $\epsilon \to 0$ where it is not important.

Our amplitude $\mathcal{M}$ of Eq.~(\ref{eq:forward-scattering-compton}) considered as an analytic
function of the complex variable $q_0$ has a branch cut along the real axis.
In order to extract the imaginary part of the diagram
we will compute the imaginary parts of the residues
by using the identity~(\ref{im-part}). We obtain
\begin{subequations}
\begin{align}
\mathrm{Im}(\mathrm{Res}(z_1))&= \frac{ \pi
 \delta( p_0+\omega_p+E_q)  }{4\omega_q E_q}\,, \displaybreak[0]\\[2ex]
\mathrm{Im}(\mathrm{Res}(z_2))&= \frac{ \pi
\delta( p_0-\omega_p-E_q)  }{4\omega_q E_q}\,, \displaybreak[0]\\[2ex]
\mathrm{Im}(\mathrm{Res}(z_3))&= \frac{ \pi
 \delta( p_0+\Omega_p+E_q)  }{4\Omega_q E_q}\,, \displaybreak[0]\\[2ex]
 \mathrm{Im}(\mathrm{Res}(z_4))&= \frac{ \pi
\delta( p_0-\Omega_p-E_q)  }{4\Omega_q E_q}\,.
\end{align}
\end{subequations}
We can then write the imaginary parts of the amplitudes as
\begin{widetext}
\begin{eqnarray}\label{Im_photon}
\mathrm{Im}(\mathcal{M}^{(1)})&=&- e^4 J_1 ^{(\lambda)}  (p',k)
 \left(\frac{\slashed{p}+m}{p^2-m^2}
\right)
   \gamma^\mu
  \int \frac{\mathrm{d}^2\vec q}{(2\pi)^3}
(\slashed{p}-\slashed{q}+m)   T_{\mu \nu} (q)
 \\
&\phantom{{}={}}& \times  \frac{(2\pi ) \pi  }{4\omega_q E_q}
\left[\delta( p_0-\omega_p-E_q) +\delta( p_0+\omega_p+E_q)   \right]
    \gamma^\nu \left(\frac{\slashed{p}+m}{p^2-m^2}
\right) J_2^{(\lambda)}   (p',k) \nonumber  \,,
\end{eqnarray}
and in the same way
\begin{eqnarray}\label{Im_ghost}
\mathrm{Im}(\mathcal{M}^{(2)})&=& e^4 J_1^{(\lambda)}   (p',k)
 \left(\frac{\slashed{p}+m}{p^2-m^2}
\right)
 \gamma^\mu
  \int \frac{\mathrm{d}^2\vec q}{(2\pi)^3}
(\slashed{p}-\slashed{q}+m   )   T_{\mu \nu} (q)
 \\
&\phantom{{}={}}& \times  \frac{(2\pi ) \pi  }{4\Omega_q E_q}
 \left[\delta( p_0-\Omega_p-E_q)+\delta( p_0+\Omega_p+E_q) \right]
    \gamma^\nu \left(\frac{\slashed{p}+m}{p^2-m^2}
\right)  J_2 ^{(\lambda)}  (p',k) \nonumber \,.
\end{eqnarray}
Now, we define
\begin{subequations}
\begin{eqnarray}
q&=&k_1  \,, \\
p-q&=&k_2\,,
\end{eqnarray}
\end{subequations}
and use energy conservation expressed by the $\delta$ functions
$\delta( p_0\pm \Omega_p \pm E_q)$
and $\delta( p_0\pm \omega_p \pm E_q)$. Furthermore, we employ the relation
\begin{equation}
\int \frac{\mathrm{d}^2\vec  q}{(2\pi)^3}=  \int \frac{\mathrm{d}^2\vec  k_1}
{(2\pi)^3}\int \frac{\mathrm{d}^2\vec  k_2}{(2\pi)^3}   \,(2\pi)^3  \delta^{(2)}  (\vec{p}-\vec{k}_1-\vec{k}_2)\,,
\end{equation}
to write the integrals over the spatial momentum components as integrals over three-momenta:
\begin{align}
2\mathrm{Im}(\mathcal{M}^{(1)})&=-e^4 J_1^{(\lambda)}(p',k)
\left(\frac{\slashed{p}+m}{p^2-m^2}\right)
\gamma^\mu \left\{\int \frac{\mathrm{d}^3 k_1}
{(2\pi)^3}\int \frac{\mathrm{d}^3 k_2}{(2\pi)^3}    \,
(\slashed{k}_2+m) \right. T_{\mu \nu} (k_1)\left[\frac{2\pi
\delta(k_1^0-\omega_{k_1}) 2\pi
\delta(k_2^0 -E_{k_2})}{(2\omega_{k_1})(2E_{k_2})}  \nonumber
  \right. \\
&\phantom{{}={}}\hspace{1cm}\left.\left.+\frac{2\pi
\delta(  k_1^0 +\omega_{k_1}) 2\pi
 \delta(  k_2^0 +E_{k_2})  }{ (2\omega_{k_1})
  (2E_{k_2})  }  \right] (2\pi)^3  \delta^{(3)}(p-k_1-k_2)\right\}  \gamma^\nu \left(\frac{\slashed{p}+m}{p^2-m^2}
\right) J_2 ^{(\lambda)}  (p',k)  \,,
\end{align}
and
\begin{align}
2\mathrm{Im}(\mathcal{M}^{(2)})&=e^4 J_1 ^{(\lambda)}  (p',k)
 \left(\frac{\slashed{p}+m}{p^2-m^2}
\right)
\gamma^\mu \left\{   \int \frac{\mathrm{d}^3 k_1}
{(2\pi)^3}\int \frac{\mathrm{d}^3 k_2}{(2\pi)^3}    \,
(\slashed{k}_2+m) \right. T_{\mu \nu} (k_1)  \left[ \frac{2\pi
\delta( k_1^0 -\Omega_{k_1})
2\pi \delta( k_2^0 -E_{k_2})  }{ (2\Omega_{k_1})
  (2E_{k_2})  }  \nonumber
  \right. \\
&\phantom{{}={}}\hspace{1cm}\left.\left.+\frac{2\pi
\delta( k_1^0 +\Omega_{k_1})
2\pi \delta(  k_2^0+E_{k_2})  }{ (2\Omega_{k_1})
  (2E_{k_2})  }  \right] (2\pi)^3  \delta^{(3)}(p-k_1-k_2)\right\}  \gamma^\nu \left(\frac{\slashed{p}+m}{p^2-m^2}
\right) J_2 ^{(\lambda)}  (p',k)  \,.
\end{align}
Recall the relations (\ref{tens-T}), (\ref{tens-T-ghost}) for
the gauge polarization vectors. Furthermore, we apply the
completeness relation (\ref{eq:completeness-particle-spinors}) for standard particle
spinors in $(2+1)$ dimensions to this particular case, i.e.,
\begin{eqnarray}
\sum_su^{(s)}(k_2)\bar{u}^{(s)}(k_2)=\slashed{k}_2+m\,,
\end{eqnarray}
where the sum runs over the spin projection $s$ of the fermion in the former loop.
Note that this spinor index is kept explicitly.
We can then write
\begin{align}
2 \mathrm{Im}(\mathcal{M}^{(1)})&=- \sum_{s,\lambda',  \lambda''} \int \frac{\mathrm{d}^3 k_1}
{(2\pi)^3}\int \frac{\mathrm{d}^3 k_2}{(2\pi)^3}
\bigg\{    -   \mathrm{i}e^2 J_1^{(\lambda)} (p',k)    \left(\frac{\slashed{p}+m}{p^2-m^2}
\right)
\gamma^\mu  u^{(s)}(k_2)      \bar e_{\mu}^{(\lambda')}  (k_1)  \bigg\}          \nonumber
   \notag \\
&\phantom{{}={}}\times g_{\lambda' \lambda''}   \bigg\{  \mathrm{i}e^2 \bar e_{\nu}^{(\lambda'')*} (k_1)
   \bar{u}^{(s)}(k_2)  \gamma^\nu  \left(\frac{\slashed{p}+m}{p^2-m^2}
\right)  J_2^{(\lambda)}  (p',k)    \bigg\}     2\pi \delta(  k^2_{1} ) 2\pi \delta(  k_{2} ^2-m^2)(2\pi)^3 \delta^{(3)}(p-k_1-k_2)
 \notag \\
&\phantom{{}={}}\times   \left[\theta (k_1^0 )
  \theta (k_2^0 )  +  \theta (-k_1^0)
  \theta (-k_2^0 ) \right] \,,
\end{align}
and
\begin{align}
2 \mathrm{Im}(\mathcal{M}^{(2)})&= \sum_{s} \int \frac{\mathrm{d}^3 k_1}
{(2\pi)^3}\int \frac{\mathrm{d}^3 k_2}{(2\pi)^3}
\bigg\{      \mathrm{i}e^2 J_1^{(\lambda)}  (p',k)    \left(\frac{\slashed{p}+m}{p^2-m^2}
\right)
\gamma^\mu u^{(s)}(k_2)     \bar  \varepsilon_{\mu}^{(+)}  (k_1)       \bigg\}      \nonumber
   \notag \\
&\phantom{{}={}}\times  \bigg\{  \mathrm{i}e^2    \bar  \varepsilon_{\nu}^{(-)} (k_1)
  \bar{u}^{(s)}(k_2)  \gamma^\nu  \left(\frac{\slashed{p}+m}{p^2-m^2}
\right)  J_2 ^{(\lambda)} (p',k)    \bigg\}  2\pi \delta(  k^2_{1} -M^2) 2\pi \delta(  k_{2} ^2-m^2)(2\pi)^3 \delta^{(3)}(p-k_1-k_2)
 \notag \\
&\phantom{{}={}}\times \left[\theta (k_1^0 )
  \theta (k_2^0 )  +  \theta (-k_1^0)
  \theta (-k_2^0 ) \right]\,.
\end{align}
In this way we obtain
\begin{align}
2 \mathrm{Im}(\mathcal{M}^{(1)})&=- \sum_{s,\lambda',  \lambda''} \int \frac{\mathrm{d}^3 k_1}
{(2\pi)^3}\int \frac{\mathrm{d}^3 k_2}{(2\pi)^3}
\bigg\{    -   \mathrm{i}e^2 J_1 ^{(\lambda)}  (p',k)    \left(\frac{\slashed{p}+m}{p^2-m^2}
\right) J_2^{(\lambda')}  (k_2,k_1)    \bigg\}        \nonumber
   \notag \\
&\phantom{{}={}}\times g_{\lambda' \lambda''} \bigg\{
  \mathrm{i}e^2 J_1^{(\lambda'')}  (k_2,k_1)   \left(\frac{\slashed{p}+m}{p^2-m^2}
\right)  J_2 ^{(\lambda)} (p',k)    \bigg\}  2\pi \delta(  k^2_{1}-M^2 ) 2\pi \delta(  k_{2} ^2-m^2) (2\pi)^3 \delta^{(3)}(p-k_1-k_2)
 \notag \\
&\phantom{{}={}}\times  \left[\theta (k_1^0 )
  \theta (k_2^0 )  +  \theta (-k_1^0)
  \theta (-k_2^0 ) \right] \,,
\end{align}
and
\begin{align}
2 \mathrm{Im}(\mathcal{M}^{(2)})&= \sum_{s} \int \frac{\mathrm{d}^3 k_1}
{(2\pi)^3}\int \frac{\mathrm{d}^3 k_2}{(2\pi)^3}
\bigg\{     \mathrm{i}e^2 J_1^{(\lambda)}  (p',k)    \left(\frac{\slashed{p}+m}{p^2-m^2}
\right) J_2^{(+)}  (k_2,k_1)    \bigg\}        \nonumber
   \notag \\
&\phantom{{}={}}\times \bigg\{   \mathrm{i}e^2 J_1^{(-)}  (k_2,k_1)
 \left(\frac{\slashed{p}+m}{p^2-m^2}
\right)  J_2 ^{(\lambda)} (p',k)    \bigg\} 2\pi \delta(  k^2_{1}-M^2 ) 2\pi \delta(  k_{2} ^2-m^2) (2\pi)^3 \delta^{(3)}(p-k_1-k_2)
 \notag \\
&\phantom{{}={}}\times \left[\theta (k_1^0 )
  \theta (k_2^0 )  +  \theta (-k_1^0)
  \theta (-k_2^0 ) \right] \,,
\end{align}
where
\begin{subequations}
\begin{align}
J_2^{(+)}  (k_2,k_1)&=\gamma^\mu u^{(s)}(k_2)
  \bar  \varepsilon_{\mu}^{(+)}(k_1)\,, \displaybreak[0]\\[2ex]
J_1^{(-)}  (k_2,k_1)&=  \bar  \varepsilon_{\nu}^{(-)} (k_1) \bar{u}^{(s)}(k_2)  \gamma^\nu\,.
\end{align}
\end{subequations}
Let us define
\begin{align}
g_{\lambda' \lambda''} \mathcal { M}_{1}^{(\lambda')} \mathcal { M}_{1}^{(\lambda'')\dag}=g_{\lambda' \lambda''}   \left(  e^2 J_1^{(\lambda)}  (p',k)
   \left(\frac{\slashed{p}+m}{p^2-m^2}
\right)
J_2^{(\lambda')}  (k_2,k_1) \right) \left(e^2 J_1^{(\lambda'')}  (k_2,k_1)   \left(\frac{\slashed{p}+m}{p^2-m^2}
\right)  J_2 ^{(\lambda)} (p',k) \right)  \,,
\end{align}
and
\begin{align}
 { \mathcal { M}^{(+)}_2}   { \mathcal { M}_2^{(-)\dag}}=    \left(e^2 J_1^{(\lambda)}  (p',k)
  \left(\frac{\slashed{p}+m}{p^2-m^2}
\right)
J_2^{(+)}  (k_2,k_1) \right)  \left(  e^2 J_1^{(-)}  (k_2,k_1)
 \left(\frac{\slashed{p}+m}{p^2-m^2}
\right)  J_2 ^{(\lambda)} (p',k)   \right)    \,.
\end{align}
%...................................................Fig5......................
\begin{figure}[t]
\centering
\includegraphics[width=0.5\textwidth]{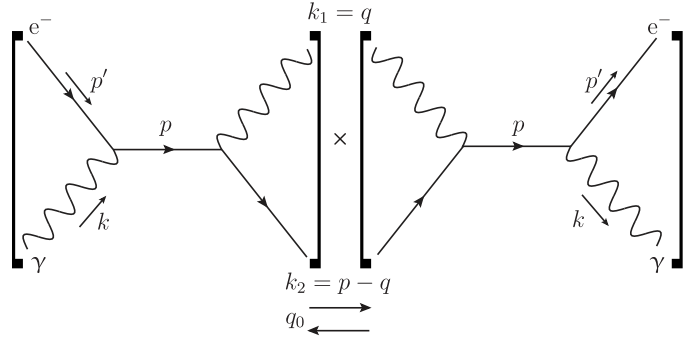}
\caption{\label{Fig5} Sum over intermediate states and energy flow in the cut Compton diagram of Fig.~\ref{Fig4}.}
\end{figure}%%
%----------------------------------------------------------------
Thus, we can express both imaginary parts as
\begin{align}
\label{eq:imaginary-part-one-loop-photon}
2 \mathrm{Im}(\mathcal{M}^{(1)})&=- \sum_{s,\lambda',\lambda''} \int \frac{\mathrm{d}^3 k_1}
{(2\pi)^3}\int \frac{\mathrm{d}^3 k_2}{(2\pi)^3}
g_{\lambda' \lambda''} \mathcal { M}_{1}^{(\lambda')} \mathcal { M}_{1}^{(\lambda'')\dag}     2\pi \delta(  k^2_{1} ) 2\pi \delta(  k_{2} ^2-m^2)
 \notag \\
&\phantom{{}={}}\times    (2\pi)^3 \delta^{(3)}(p-k_1-k_2) \left[\theta (k_1^0 )
  \theta (k_2^0 )  +  \theta (-k_1^0)
  \theta (-k_2^0 ) \right] \,,
\end{align}
and
\begin{align}
\label{eq:imaginary-part-one-loop-ghost}
2 \mathrm{Im}(\mathcal{M}^{(2)})&= -\sum_{s} \int \frac{\mathrm{d}^3 k_1}
{(2\pi)^3}\int \frac{\mathrm{d}^3 k_2}{(2\pi)^3}
 { \mathcal { M}^{(+)}_2}   { \mathcal { M}_2^{(-)\dag}}    2\pi \delta(  k^2_{1}-M^2 ) 2\pi \delta(  k_{2} ^2-m^2)
 \notag \\
&\phantom{{}={}}\times    (2\pi)^3 \delta^{(3)}(p-k_1-k_2) \left[\theta (k_1^0 )
  \theta (k_2^0 )  +  \theta (-k_1^0)
  \theta (-k_2^0 ) \right]\,.
\end{align}
The sum in~Eq.~(\ref{eq:imaginary-part-one-loop-photon}) runs over the spin projection
of the fermion and the polarization of the photon. Both particles
were put on-shell by cutting the diagram of the forward-scattering amplitude (see~Fig.~\ref{Fig5})
into two pieces. Note that the right-hand side of Eq.~(\ref{eq:imaginary-part-one-loop-ghost}) is zero,
as the $\delta$ function does not provide a contribution due to the large mass scale $M$.
This behavior is precisely the effect of the Lee-Wick prescription according
to which the negative-norm states are removed from the asymptotic Hilbert space just as in the tree-level analysis
(see Eq.~(\ref{final_result}) and the subsequent paragraph). So, we conclude that the optical theorem and, therefore,
unitarity continue being valid at one-loop order, as well.
\end{widetext}

%%%%%%%%%%%%%%%%%%%%%%%%%%
\section{Conclusions and outlook}\label{sec5}
%%%%%%%%%%%%%%%%%%%%%%%%%%
In this paper, we considered a higher-derivative Chern-Simons-type modification
of electrodynamics in $(2+1)$ dimensions.
We decomposed the Lagrangian of the model into a physical and a ghost sector and
obtained the polarization vectors for the
corresponding modes. In addition, the propagator of the theory was computed and it was
demonstrated how it can be expressed in terms of
the polarization vectors. Based on these findings, we performed the canonical quantization
of the theory and studied its perturbative unitarity at both tree-level and one-loop order by
checking the validity of the optical theorem.

Throughout this paper, we explicitly demonstrated that reflection positivity, known as a
sufficient condition for unitarity, is satisfied. As the latter requirement
applies to a free field theory only, we were interested in understanding unitarity when
taking interactions into account. Hence, we coupled our theory to standard Dirac fermions
in $(2+1)$ spacetime dimensions and evaluated the optical theorem for particular scattering
processes. This analysis of unitarity revealed inconsistencies due to negative contributions
at the pole of the ghost, as one should expect.
However, by using the Lee-Wick prescription we have demonstrated
that unitarity is conserved at both tree-level and one-loop order. The method of
removing contributions from ghosts from the in- and out-states clearly provides this result.
We applied the usual
cutting rules of Feynman diagrams and amplitudes to guarantee the validity of the optical theorem. It was
necessary to assume that the ghost mass is high enough, perhaps of the order of the Planck mass.
It is expected that the situation at higher order in perturbation theory will not be very different.

It is also reasonable to expect that these results can be generalized naturally to the four-dimensional case
where the higher-derivative Chern-Simons-like term breaks Lorentz symmetry. Some preliminary studies of unitarity in this alternative
theory have been carried out in~\cite{Bhabha1}. They are complemented by the analysis performed in our latest work \cite{Ferreira:2020wde}.

Moreover, our opinion is that the results obtained here could serve as a base to explicitly define
classes of higher-derivative theories consistent with the requirement of unitarity. In particular,
our methodology could be useful for studies of various higher-derivative
extensions of gravity including the Lorentz-breaking ones. We hope
that this methodology will help to solve the problem of formulating a perturbatively consistent gravity model.

\begin{acknowledgments}
R.A. has been supported by project Ayudant\'{i}a de Investigaci\'{o}n No. 352/1959/2017 and 352/12361/2018 of Universidad del B\'{i}o-B\'{i}o.
The work by A.Yu.P. has been supported by CNPq Produtividade 303783/2015-0. C.M.R acknowledges
support from Fondecyt Regular project No.~1191553, Chile. M.S. is indebted to FAPEMA
Universal 01149/17, CNPq Universal 421566/2016-7,
and CNPq Produtividade 312201/2018-4.
\end{acknowledgments}

\begin{appendix}
%................................................................................
\section{Dirac formalism}\label{App:A}
%................................................................................

\setcounter{equation}{0}
\renewcommand{\theequation}{A.\arabic{equation}}

We follow the Dirac procedure to reduce second-class
constraints from the higher-derivative theory based on the Lagrangian~(\ref{L-D})
to zero~\cite{testing}
and to find the Dirac brackets. From Eqs.~(\ref{second-cons})
and~(\ref{pi}), we have four primary second-class
constraints%%
\begin{subequations}
\begin{eqnarray}
\chi_0(t,\vec x)&=&\Pi^0(t,\vec x)-\frac{g}{2} \epsilon^{ij}
\partial_iA_j (t,\vec x)\,, \\   \chi_1(t,\vec x)&=&P_0(t,\vec x)+
 \dot A_0(t,\vec x)
 +\frac{g}{2} \epsilon^{ij}
\partial_i \dot A_j (t,\vec x)\,, \\
\varphi^i (t,\vec x)&=&\Pi^i (t,\vec x)+\frac{g}{2} \epsilon^{ij}  \dot A_{j}(t,\vec x)
 - \frac{g}{2} \epsilon^{ij}  \partial_j A_{0}(t,\vec x)\,. \nonumber \\
\end{eqnarray}
\end{subequations}
The non-vanishing elements of the algebra are
\begin{subequations}
\begin{eqnarray}
\left\{  \chi_1(t,\vec x),\chi_0(t,\vec y)\right\}&=&\delta ^{(2)}
(\vec{x}-\vec y)\,,
\\[1ex]
 \left\{ \varphi^i(t,\vec x), \chi_1(t,\vec y) \right\}&=&-g\epsilon^{ij}
\partial_j \, \delta ^{(2)}(\vec{x}-\vec y) \,,
 \\[1ex] \label{CONSTRAINTALG}
\left\{ \varphi^i(t,\vec x),\varphi^j(t,\vec y)\right\}&=&g
\epsilon^{ij}\delta ^{(2)}(\vec{x}-\vec y)\,.
\end{eqnarray}
\end{subequations}
The convention we use is that the derivatives act on the first set of spatial variables
named $\vec x$, in general. To begin, let us
introduce the notation $\varphi_{A}=(\chi_0,\chi_1 ,\varphi^i )$, with $A=\bar 0,\bar 1,1,2$
and $i=1,2$.
The matrix of the second-class constraints will be denoted by
\begin{eqnarray}
C_{AB}(t;{\vec x},{\vec y})=\{
\varphi_{A}(t,{\vec x}),\varphi_{B}(t,{\vec y}) \} \,.
\end{eqnarray}
From~Eq.~(\ref{CONSTRAINTALG}) we have
\begin{equation}
C_{AB}=\left[
\begin{array}{cccc}
0 & -1 & 0 & 0  \\
1 & 0 & -g\partial_2 &  g\partial_1 \\
0& -g\partial_2 & 0 & g  \\
0& g\partial_1 & -g & 0
\end{array}
\right] \delta ^{(2)}(\vec{x}-\vec{y})\,.
\end{equation}
The inverse matrix is (where the $\delta$ function is not inverted):
\begin{equation}
C_{AB}^{-1}=\left[
\begin{array}{cccc}
0 & 1 & -\partial_1 & -\partial_2 \\
-1 & 0 & 0 &  0 \\
-\partial_1& 0 & 0 & -1/g  \\
-\partial_2& 0 & 1/g & 0
\end{array}
\right] \delta ^{(2)}(\vec{x}-\vec{y})\,.
\end{equation}
The nonzero components are
\begin{subequations}
\begin{align}
 C_{\bar 0\bar 1}^{-1}(\vec{x},\vec{y})
 &=-C_{\bar 1\bar 0}^{-1}(\vec{x},\vec{y})=\delta ^{(2)}(\vec{x}-\vec{y})\,,
\\[1ex]
 C_{\bar 0i}^{-1}(\vec{x},\vec{y})
 &=C_{i\bar 0}^{-1}(\vec{x},\vec{y})=-\partial_i\delta ^{(2)}(\vec{x}-\vec{y})\,,
\\[1ex]
 C_{ij}^{-1}(\vec{x},\vec{y})
 &=-\frac{1}{g}\, \epsilon^{ij}\delta ^{(2)}
 (\vec{x}-\vec{y}),    \qquad i,j=1,2\,.
\end{align}
\end{subequations}
The Dirac brackets are defined by
\begin{equation}
\{ X, Y\}^*=\{ X, Y\}-\{ X, \varphi_A\}\;
C^{-1}_{AB}\;
 \{\varphi_B, Y\} \,.
\end{equation}
We promote the Dirac algebra to the equal-time commutators
satisfied by the fields and obtain
\begin{subequations}
\begin{align}
\left[A_0( t,  \vec{x}  ), \dot A_0(t,\vec{y} )\right]&=-\mathrm{i} \delta^{(2)}(\vec{x}-\vec{y}) \,,
\label{Com1}\displaybreak[0]\\[1ex]
\left[ A_0({t,\vec x  }), P_0( {t,\vec{y}} )\right]&=\mathrm{i}\delta^{(2)}(\vec{x}-\vec{y}) \,,
 \label{Com2} \displaybreak[0]\\[1ex]
\left[A_0({t, \vec x}), P^i(t,{\vec{y}})\right]
&=\frac{\mathrm{i}g}{2}\epsilon^{ij}\partial_j\, \delta^{(2)}(\vec{x}-\vec{y}) \,,
\label{Com3} \displaybreak[0]\\[1ex]
\left[ \dot A_i( {t,\vec{x}}), \dot A_j(t,{\vec{y} })\right]
&=-\frac{\mathrm{i}}{g}\epsilon^{ij}\delta^{(2)}(\vec{x}-\vec{y})\,,
\label{Com4} \displaybreak[0]\\[1ex]
\left[\dot  A_i({t,\vec{x}}), \dot A_0 (t,{\vec{y}})\right]
&=-\mathrm{i}\partial_i\,\delta^{(2)}(\vec{x}-\vec{y})\,,
\label{Com5} \displaybreak[0]\\[1ex]
\left[\dot  A_i({t,\vec{x}}), P_0 (t,{\vec{y}})\right]
&=\frac{\mathrm{i}}{2}\partial_i\,\delta^{(2)}(\vec{x}-\vec{y})\,,
\label{Com6} \displaybreak[0]\\[1ex]
\left[\dot  A_i({t,\vec{x}}), P^j (t,{\vec{y}})\right]&
=\frac{\mathrm{i}g}{2}\epsilon^{jk}\partial_i\partial_k\delta^{(2)}(\vec{x}-\vec{y})\,,
\label{Com7} \displaybreak[0]\\[1ex]
\left[\dot  A_i({t,\vec{x}}), \Pi^j (t,{\vec{y}})\right]
&=\frac{\mathrm{i}}{2}\delta_{ij}\,\delta^{(2)}(\vec{x}-\vec{y})\,,
\label{Com8} \displaybreak[0]\\[1ex]
\left[P_0({t,\vec{x}}), \Pi^i (t,{\vec{y}})\right]&= \frac{\mathrm{i}g}{4}
\epsilon^{ij}\partial_j\,\delta^{(2)}(\vec{x}-\vec{y})\,,
\label{Com10} \displaybreak[0]\\[1ex]
\left[\Pi_0({t,\vec{x}}), P^i (t,{\vec{y}})\right]&= -\frac{\mathrm{i}g}{2}
\epsilon^{ij}\partial_j\,\delta^{(2)}(\vec{x}-\vec{y})\,,
\label{Com12} \displaybreak[0]\\[1ex]
\left[\Pi^i({t,\vec{x}}), \Pi^j (t,{\vec{y}})\right]
&= -\frac{\mathrm{i}g}{4}\epsilon^{ij} \delta^{(2)}(\vec{x}-\vec{y})  \label{Com13} \,.
\end{align}
\end{subequations}
Note that the momentum $P^{\mu}$
has been changed in comparison to that employed in Ref.~\cite{testing}
and, consequently, we have obtained a different algebra.
%................................................................................
\section{The Hamiltonian }\label{App:B}
%................................................................................

\setcounter{equation}{0}
   \renewcommand{\theequation}{B.\arabic{equation}}
The current section delivers a detailed demonstration on how the
Hamiltonian of the theory given by Eq.~(\ref{L-D}) can be expressed in terms of
creation and annihilation operators. We consider the Hamiltonian~(\ref{Ham}) written as
\begin{eqnarray}
H=H_{\bar{A}}+H_G \,,
\end{eqnarray}
where by using the decomposition~(\ref{planewave}) we have
\begin{eqnarray}
H_{\bar{A}}&=&\int \mathrm{d}^2x\left( - \frac 12 \dot{\bar A}_\mu ( \hat U^{\mu \nu } )  \dot{\bar A}_\nu +
  \frac{1}{2}  \bar A_\mu  (\hat  U^{\mu \nu } ) \nabla^2 \bar A_\nu    \right)\,, \nonumber \\
\\[1ex]
H_G&=&\frac{g}{2} \int \mathrm{d}^2x\,\epsilon^{ij }  \dot{G}_i \Box G_j   \,.
\end{eqnarray}
Above, we have applied the
equation of motion~(\ref{constraints-second}) for the ghost and $\Box \bar A_{\mu}=0$ for the photon.

Let us define
\begin{eqnarray}
H^{\text{kin}}_{\bar{A}}&=&- \frac 12 \int \mathrm{d}^2x\,  \dot{\bar A}_\mu  \hat U^{\mu \nu }   \dot{\bar A}_\nu  \,,
\\
H^{\text{pot}}_{\bar{A}}&=&\frac{1}{2} \int \mathrm{d}^2x\,
    \bar A_\mu  \hat  U^{\mu \nu }  \nabla^2 \bar A_\nu    \,.
\end{eqnarray}
Inserting the photon field operator of~Eq.~(\ref{photon-field}), the first contribution reads

\begin{widetext}

\begin{align}
H^{\text{kin}}_{\bar{A}}&= \frac 18  \int\frac{\mathrm{d}^2 \vec p }{(2\pi)^2}
 \sum_{\lambda,\lambda'} \left[a^{( \lambda)}_{\vec p} a^{( \lambda')}_{-\vec p}
  \bar{e}^{(\lambda)}_{\mu} (\vec p)
U^{ \mu \nu } _{-\vec p} \, \bar{e}^{(\lambda')}_{\nu}(-\vec p)
  e^{-2\mathrm{i} \omega_{\vec p}  x_0}   -a^{( \lambda)}_{\vec p}
  a^{( \lambda')\dag }_{ \vec p}  \bar{e}^{(\lambda)}_{\mu} (\vec p)
U^{ \mu \nu *}  _{\vec p}\bar{e}^{(\lambda')*}_{\nu}(\vec p)
    \right. \nonumber \\[1ex]
&\phantom{{}={}}\hspace{2.4cm} \left. -\,a^{( \lambda)\dag}_{\vec p} a^{( \lambda') }_{\vec p}  \bar{e}^{(\lambda)*}_{\mu} (\vec p)
U^{ \mu \nu } _{\vec p} \bar{e}^{(\lambda')}_{\nu}(\vec p)
 +a^{( \lambda)\dag }_{\vec p} a^{( \lambda')\dag }_{-\vec p}  \bar{e}^{(\lambda)*}_{\mu} (\vec p)
U^{ \mu \nu *} _{-\vec p}\,  \bar{e}^{(\lambda')*}_{\nu}(-\vec p) e^{2\mathrm{i} \omega_{\vec p} x_0}\right]    \,,
\end{align}
where $U^{\mu \nu }_{\vec p }=(\eta^{\mu \nu}-\mathrm{i}g\epsilon^{\mu \beta \nu}p_{\beta})_{p_0=\omega_p}$ corresponds
to~Eq.~(\ref{def_U}) in momentum space.

In the same way,
\begin{align}
H^{\text{pot}}_{\bar{A}}&=- \frac 18   \int\frac{\mathrm{d}^2 \vec p }{(2\pi)^2}
\sum_{\lambda,\lambda'}   \frac{\vec p^{\,2}}{\omega_{\vec p}^2} \left[a^{( \lambda)}_{\vec p}
a^{( \lambda')}_{-\vec p}  \bar{e}^{(\lambda)}_{\mu} (\vec p)
U^{ \mu \nu } _{-\vec p} \, \bar{e}^{(\lambda')}_{\nu}(-\vec p)
  e^{-2\mathrm{i} \omega_{\vec p}  x_0}    +a^{( \lambda)}_{\vec p}
  a^{( \lambda')\dag }_{ \vec p}  \bar{e}^{(\lambda)}_{\mu} (\vec p)
U^{ \mu \nu *}  _{\vec p}\bar{e}^{(\lambda')*}_{\nu}(\vec p)
     \right. \nonumber \\[1ex]
&\phantom{{}={}}\hspace{3.2cm}\left.  +\,a^{( \lambda)\dag}_{\vec p} a^{( \lambda') }_{\vec p}  \bar{e}^{(\lambda)*}_{\mu} (\vec p)
U^{ \mu \nu } _{\vec p} \bar{e}^{(\lambda')}_{\nu}(\vec p)  +
 a^{( \lambda)\dag }_{\vec p} a^{( \lambda')\dag }_{-\vec p}  \bar{e}^{(\lambda)*}_{\mu} (\vec p)
U^{ \mu \nu *} _{-\vec p}\,  \bar{e}^{(\lambda')*}_{\nu}(-\vec p) e^{2\mathrm{i} \omega_{\vec p} x_0}\right]     \,,
\end{align}

\end{widetext}
We see that the first and last terms vanish due to the
global factor $1-\vec p^{\,2}/\omega_p^2$, while the other terms pick up a factor of
$1+\vec p^{\,2}/ \omega_p^2=2$.
We arrive at
\begin{eqnarray}
H_{\bar{A}}=- \frac 14   \int\frac{\mathrm{d}^2 \vec p }{(2\pi)^2}
\sum_{\lambda,\lambda'}   \eta_{\lambda \lambda'} \Big(  a^{( \lambda)}_{\vec p} a^{( \lambda')\dag }_{ \vec p}
   + a^{( \lambda)\dag}_{\vec p} a^{( \lambda') }_{\vec p}    \Big)    \,, \nonumber \\
\end{eqnarray}
where we have used
\begin{eqnarray}
 \bar e^{(\lambda)}_{\mu}   U^{ \mu \nu * }_{\vec p}\bar e^{*(\lambda')} _{\nu} =g^{\lambda \lambda'}   \,.
\end{eqnarray}
and its complex conjugated.

For the ghost part we insert the ghost field operator of~Eq.~(\ref{ghost-field}) and obtain
\begin{align}
H_G&=\frac g8   \int\frac{\mathrm{d}^2 \vec p }{(2\pi)^2}
\frac{\mathrm{i} \epsilon^{ij } }{g^2 \Omega_p} \left[  b_{\vec p} \,
b^{\dag} _{\vec p}\,\bar\varepsilon_i^{(+)}( \vec p)\,\bar\varepsilon_j^{(+)*}( \vec p)  \right. \nonumber \\
&\phantom{{}={}}\hspace{2cm}\left.-b^{\dag}_{\vec p} \, b _{\vec p} \,  \bar \varepsilon_i^{(+)*}( \vec p)\,  \bar \varepsilon_j^{(+)}( \vec p)    \right]  \,,
\end{align}
where we have used that $p^2=1/g^2$ as well as
\begin{subequations}
\begin{equation}
\epsilon^{ij}   \bar \varepsilon_i^{(+)}( \vec p)   \bar \varepsilon_j^{(+)}( -\vec p) =0\,,
\end{equation}
since
\begin{equation}
\bar \varepsilon_k^{(+)}( -\vec p) =-  \bar \varepsilon_k^{(+)}( \vec p)\,,
\end{equation}
\end{subequations}
for $k=1,2$. We then arrive at
\begin{eqnarray}
H_G&=& -\frac 14   \int\frac{\mathrm{d}^2 \vec p }{(2\pi)^2}
 \left(  b_{\vec p} \, b^{\dag} _{\vec p}
 + b^{\dag}_{\vec p} \, b _{\vec p}    \right)   \,,
\end{eqnarray}
where we have also employed
\begin{equation}
\epsilon^{ij } \bar  \varepsilon_i^{(\pm)}  \bar \varepsilon_j^{(\pm)\ast}|_{p_0=\Omega_p} =\pm 2 \mathrm{i}g \Omega_p  \,.
\end{equation}
This proves our expression~(\ref{Ham2}).
%...................................................................................
\section{Extended equal-time commutators}\label{App:C}
%...................................................................................
\setcounter{equation}{0}
   \renewcommand{\theequation}{C.\arabic{equation}}
In this section we intend to compute the equal-time commutators for the field operators that
emerge from field theory of higher derivatives defined by Eq.~(\ref{L-D}).
Consider the basic commutator
\begin{subequations}
\begin{align} \label{basic}
&\left[  A_{\mu}(x),   A_{\nu}(y)  \right]=\left[  \bar A_{\mu}(x),   \bar A_{\nu}(y)  \right] +
\left[  G_{\mu}(x),   G_{\nu}(y)  \right] \,.
\end{align}
with
\begin{align}
\left[\bar A_{\mu}(x),\bar A_{\nu}(y)\right]&=-\int \frac{\mathrm{d}^2\vec p}
{(2\pi)^2}\frac{1}{2\omega_p}\left( T_{\mu\nu}(p) e^{-\mathrm{i}p\cdot (x-y)}
 \right. \nonumber \\
&\phantom{{}={}}\left.-T_{\nu\mu}(p) e^{\mathrm{i}p\cdot (x-y)}\right)_{p_0=\omega_p}\,,
\displaybreak[0]\\[2ex]
\left[G_{\mu}(x),G_{\nu}(y)\right]&=\int \frac{\mathrm{d}^2\vec p}
{(2\pi)^2}\frac{1}{2\Omega_p}\left(T_{\mu\nu}(p)
e^{-\mathrm{i}p\cdot (x-y)}    \right. \nonumber \\
&\phantom{{}={}}\left.-T_{\nu\mu} (p)e^{\mathrm{i}p\cdot (x-y)}\right)_{p_0=\Omega_p}\,.
\end{align}
\end{subequations}
To derive the Dirac commutators we work directly with the field operators of Eqs.~(\ref{photon-field})
and~(\ref{ghost-field}). Our strategy will be as follows:
\begin{itemize}
\item[a)] We consider the basic commutator~(\ref{basic})
and construct the various elements in phase space by
applying the different operators on the fields.
\item[b)] For a commutator containing $\Box A_{\mu} (t,{\vec {x}} )$ we use the identities
 $\Box \bar A_{\mu}(t,{\vec {x}} )=0$ and $\Box G_{\mu}(t,{\vec {x}} )=-\frac{1}{g^2}
G_{\mu}(t,{\vec {x}} )$.
\item[c)] Whenever an integral involves momentum variables
we use the relation $p_{\mu}=\mathrm{i}\partial_{\mu}$, whereupon derivatives can be extracted from
the integral.
\item[d)] To treat derivatives for the second variable $\partial_{i}^ y$, we integrate
 by parts to produce $\partial_{i}^ x$ whereby an additional minus sign occurs.
\item[e)] We assume that the spatial derivatives $\partial_i$ act
  on the first variable $\vec x$ of $\delta$ functions in all final expressions.
\end{itemize}
%............................................................................................................
\addtocontents{toc}{\protect\setcounter{tocdepth}{1}}
\subsection{Commutator $\boldsymbol{[A_0(t,\vec x),\dot{{A}}_0 (t,\vec y)]}$}\label{SubC1}
\addtocontents{toc}{\protect\setcounter{tocdepth}{2}}
%............................................................................................................
With the previous rules in mind and to demonstrate our technique explicitly
we apply a first time derivative $\partial_{y_0}$ to the basic
commutator (\ref{basic}):
\begin{align}
&\left[   A_{\mu}(x),   \partial_{y_0} A_{\nu}(y)  \right] \nonumber \\ &=-\int \frac{\mathrm{d}^2\vec p}
{(2\pi)^2}    \frac{ \mathrm{i} }{2 } \left( T_{\mu\nu} e^{-\mathrm{i}p\cdot (x-y)}
 +T_{\nu\mu} e^{\mathrm{i}p\cdot (x-y)} \right) _{p_0=\omega_p}
\nonumber \\
&\phantom{{}={}}+\int \frac{\mathrm{d}^2\vec p}
{(2\pi)^2}    \frac{\mathrm{i} }{2  } \left( T_{\mu\nu}
e^{-\mathrm{i}p\cdot (x-y)} +T_{\nu\mu} e^{\mathrm{i}p\cdot (x-y)}  \right)_{p_0=\Omega_p} \,.
\end{align}
We set both times equal, $x_0=y_0=t$, and change $\vec p \to -\vec p$
in the second term of each contribution. We then obtain
\begin{align}
&\left[   A_{\mu}(t, \vec x),   \dot A_{\nu}(t, \vec y)  \right]\nonumber \\ &=-\int \frac{\mathrm{d}^2\vec p}
{(2\pi)^2}    \frac{\mathrm{i}}{2 } \left( T_{\mu\nu} (\vec p)
 +T_{\nu\mu}  (-\vec p)  \right) e^{\mathrm{i} \vec p\cdot (\vec x-\vec y)}
\nonumber \\
&\phantom{{}={}}+\int \frac{\mathrm{d}^2\vec p}
{(2\pi)^2}    \frac{\mathrm{i} }{2  } \left( T_{\mu\nu}  (\vec p)
 +T_{\nu\mu}  (-\vec p) \right)  e^{\mathrm{i} \vec p\cdot (\vec x-\vec y)}  \,.
\end{align}
In the following calculations we implicitly consider the dependence on $\omega_p$
and  $\Omega_p$ of the expressions in parentheses above.
For the indices $\mu=0$ and $\nu=0$, we have
\begin{align}
\left[   A_{0}(t, \vec x),   \dot A_{0}(t, \vec y)  \right]&=-\int \frac{\mathrm{d}^2\vec p}
{(2\pi)^2}    \frac{ \mathrm{i} }{2 } \left( 2-2g^2\omega_p^2 \right) e^{\mathrm{i} \vec p\cdot (\vec x-\vec y)}
\nonumber \\
&\phantom{{}={}}+\int \frac{\mathrm{d}^2\vec p}
{(2\pi)^2}    \frac{\mathrm{i} }{2  } \left( 2-2g^2\Omega_p^2  \right) e^{\mathrm{i} \vec p\cdot (\vec x-\vec y)}  \,,
\end{align}
where we have used $T_{00}(\vec p)+T_{00}(-\vec p)=2-2g^2p_0$. Adding both terms yields
\begin{align}
\left[   A_{0}(t, \vec x),   \dot A_{0}(t, \vec y)  \right]=\int \frac{\mathrm{d}^2\vec p}
{(2\pi)^2}   \mathrm{i} g^2\left( \omega_p^2-\Omega_p^2 \right) e^{\mathrm{i} \vec p\cdot (\vec x-\vec y)}  \,,
\end{align}
and since
\begin{eqnarray}
 \omega_p^2-\Omega_p^2 =-\frac{1}{g^2}\,,
\end{eqnarray}
one arrives at the first commutator~(\ref{Com1}):
\begin{eqnarray}\label{A_0}
\left[   A_{0}(t, \vec x),   \dot A_{0}(t, \vec y)  \right]&=&-\mathrm{i} \delta^{(2)}(\vec x-\vec y) \,.
\end{eqnarray}
%............................................................................................................
\addtocontents{toc}{\protect\setcounter{tocdepth}{1}}
\subsection{Commutator $\boldsymbol{[ A_0({t,\vec{x}}), P_0( {t,\vec{y}} )]}$}\label{SubC2}
\addtocontents{toc}{\protect\setcounter{tocdepth}{2}}
%............................................................................................................
Here we compute an unmodified commutator by using our method.
Recall~Eq.~(\ref{second-cons}) and write
\begin{eqnarray}
P_0(t, \vec y)=-\dot A_0(t, \vec y)-\frac{g}{2}\epsilon^{ij}\partial_i \dot A_j(t, \vec y)\,.
\end{eqnarray}
We get
\begin{align}
\left[ A_0({t,\vec{x}}), P_0( {t,\vec{y}} )\right]&=\left[A_0({t,\vec{x}}),-\dot A_0(t,\vec y)\right. \notag \\
&\phantom{{}={}}\hspace{1cm}\left.-\frac{g}{2}\epsilon^{ij}\partial_i \dot A_j(t, \vec y) \right] \,.
\end{align}
The second commutator is zero, i.e.,
\begin{eqnarray}\label{comm3}
 \left[ A_0(t,\vec x), \dot A_j(t,{\vec {y}} )\right] =0\,,
\end{eqnarray}
and using the result~(\ref{A_0}) we arrive at
\begin{eqnarray}
\left[ A_0({t,\vec{x}}), P_0( {t,\vec{y}} )\right]&=&\mathrm{i} \delta^{(2)}(\vec x-\vec y)\,,
\end{eqnarray}
which gives~Eq.~(\ref{Com2}).
%............................................................................................................
\addtocontents{toc}{\protect\setcounter{tocdepth}{1}}
\subsection{Commutator $\boldsymbol{[A_0({t,\vec {x}}), P^i(t,{\vec {y}} )]}$}\label{SubC3}
\addtocontents{toc}{\protect\setcounter{tocdepth}{2}}
%............................................................................................................
It follows from (\ref{second-cons}) that the spatial momentum components read
\begin{align}\label{space-momenta}
P^i=-\dot A^i+\frac{g}{2}  \epsilon^{ik}\Box  A_k+\frac{g}{2}  \epsilon^{ik}
 \ddot A_k-\frac{g}{2}  \epsilon^{ik} \partial_k \dot A_0\,.
\end{align}
Then
\begin{align}
\left[A_0({t,\vec {x}}), P^i(t,{\vec {y}} )\right]&=\left[ A_0(t,\vec x),
 -\dot A^i(t,{\vec {y}} )+\frac{g}{2}  \epsilon^{ik}\Box  A_k
(t,{\vec {y}} )\nonumber \right.  \\
&\phantom{{}={}}\,\left.+\,\frac{g}{2}  \epsilon^{ik}
 \ddot A_k (t,{\vec {y}} )-\frac{g}{2}  \epsilon^{ik} \partial_k
  \dot A_0 (t,{\vec {y}} )\right]\,.
\end{align}
We take into account that the first commutator is zero; see Eq.~(\ref{comm3}).
Furthermore, we employ $\Box \bar A_k(t,{\vec {y}} )=0$ and $\Box G_k(t,{\vec {y}} )=-\frac{1}{g^2}
G_j(t,{\vec {y}} )$ in the second to arrive at
\begin{align}
\label{thecomm}
\left[A_0({t,\vec {x}}), P^i(t,{\vec {y}} )\right]&=-\frac{1}{2g}  \epsilon^{ik}
  \left[ G_0(t,\vec x),   G_k
(t,{\vec {y}} ) \right] \nonumber \\
&\phantom{{}={}}  + \frac{g}{2}  \epsilon^{ik} \left[  A_0({t,\vec {x}}) ,
\ddot A_k (t,{\vec {y}} )  \right]
\nonumber \\
&\phantom{{}={}} +\frac{g}{2}  \epsilon^{ik} \partial_k   \left[ A_0({t,\vec {x}}),
 \dot A_0 (t,{\vec {y}} )\right]\,,
\end{align}
where the final spatial derivative has been integrated by parts.

One can show that
\begin{eqnarray}
  \left[ G_0(t,\vec x),   G_k
(t,{\vec {y}} ) \right]  =-\mathrm{i}g^2\partial _k \delta^{(2)}(\vec x-\vec y)\,,
\end{eqnarray}
and also
\begin{eqnarray}
 \left[  A_0({t,\vec {x}}) ,\ddot A_k (t,{\vec {y}} )  \right]=\mathrm{i} \partial _k \delta^{(2)}(\vec x-\vec y)\,.
\end{eqnarray}
Substituting these expressions in~Eq.~(\ref{thecomm}), we obtain
\begin{align}
\left[A_0({t,\vec {x}}), P^i(t,{\vec {y}} )\right]&=- \frac{1}{2g}  \epsilon^{ik}
\left (- \mathrm{i}g^2\partial _k \delta^{(2)}(\vec x-\vec y) \right) \nonumber \\
&\phantom{{}={}}+ \frac{g}{2}  \epsilon^{ik}
 \left (\mathrm{i} \partial _k \delta^{(2)}(\vec x-\vec y) \right)
\nonumber \\
&\phantom{{}={}}+\frac{g}{2}  \epsilon^{ik} \partial_k
  \left(-\mathrm{i} \delta^{(2)}(\vec x-\vec y) \right)\,.
\end{align}
The last two terms cancel and we arrive at
\begin{eqnarray}
\left[A_0({t,\vec {x}}), P^i(t,{\vec {y}} )\right]&=& \frac{\mathrm{i}g}{2}
  \epsilon^{ik} \partial_k  \delta^{(2)}(\vec x-\vec y) \,,
\end{eqnarray}
which is~Eq.~(\ref{Com3}).
%............................................................................................................
\addtocontents{toc}{\protect\setcounter{tocdepth}{1}}
\subsection{Commutator $\boldsymbol{[\dot A_i({t,\vec {x}}), \dot A_j (t,{\vec {y}} )]}$}\label{SubC4}
\addtocontents{toc}{\protect\setcounter{tocdepth}{2}}
%............................................................................................................
To derive~Eq.~(\ref{Com4}), it follows from~Eq.~(\ref{basic}) that
\begin{align}
&\left[  \dot  A_{i}(t, \vec x),   \dot A_{j}(t, \vec y)  \right] \nonumber \\ &=-\int \frac{\mathrm{d}^2\vec p}
{(2\pi)^2}    \frac{ \omega_p }{2 } \left( T_{ij} (\vec p)
 -T_{ji}  (-\vec p)  \right) e^{\mathrm{i} \vec p\cdot (\vec x-\vec y)}
\nonumber \\
&\phantom{{}={}}+\int \frac{\mathrm{d}^2\vec p}
{(2\pi)^2}    \frac{\Omega_p }{2  } \left( T_{ij}  (\vec p)
 -T_{ji}  (-\vec p) \right)  e^{\mathrm{i} \vec p\cdot (\vec x-\vec y)}  \,,
\end{align}
and applying the definition (\ref{eq:tensor-T}) we find
\begin{eqnarray}
T_{ij} (\vec p)
 -T_{ji}  (-\vec p)  =-2\mathrm{i}g\epsilon_{ij} p_0\,.
\end{eqnarray}
Thus, we have
\begin{align}
\left[  \dot  A_{i}(t, \vec x),   \dot A_{j}(t, \vec y)  \right]&=-\int \frac{\mathrm{d}^2\vec p}
{(2\pi)^2}   (  -\mathrm{i}g\epsilon_{ij} \omega_p^2)  e^{\mathrm{i} \vec p\cdot (\vec x-\vec y)}
 \nonumber\\
&\phantom{{}={}}+\int \frac{\mathrm{d}^2\vec p}
{(2\pi)^2}  (  -\mathrm{i}g\epsilon_{ij} \Omega_p^2) e^{\mathrm{i} \vec p\cdot (\vec x-\vec y)}  \nonumber\\
&=\mathrm{i}g\epsilon_{ij}
\int \frac{\mathrm{d}^2\vec p}
{(2\pi)^2}   (\omega_p^2-\Omega_p^2) e^{\mathrm{i} \vec p\cdot (\vec x-\vec y)}   \,,
\end{align}
and finally,
\begin{equation}
\label{comm4}
\left[  \dot  A_{i}(t, \vec x),   \dot A_{j}(t, \vec y)  \right]=-\frac{\mathrm{i}}{g} \epsilon_{ij} \delta^{(2)}(\vec x-\vec y) \,.
\end{equation}
%............................................................................................................
\addtocontents{toc}{\protect\setcounter{tocdepth}{1}}
\subsection{Commutator $\boldsymbol{[\dot A_i({t,\vec {x}}), \dot A_0 (t,{\vec {y}} )]}$}\label{SubC5}
\addtocontents{toc}{\protect\setcounter{tocdepth}{2}}
%............................................................................................................
Repeating the calculations performed in subsection~\ref{SubC1} we find
\begin{align}
&\left[\dot A_i({t,\vec {x}}), \dot A_0 (t,{\vec {y}} )\right] \nonumber  \\ &=- \int \frac{\mathrm{d}^2\vec p}
{(2\pi)^2}    \frac{ \omega_p }{2 } \left( T_{i0} (\vec p)
 -T_{0i}  (-\vec p)  \right) e^{\mathrm{i} \vec p\cdot (\vec x-\vec y)}
\nonumber \\
&\phantom{{}={}}+ \int \frac{\mathrm{d}^2\vec p}
{(2\pi)^2}    \frac{\Omega_p }{2  } \left( T_{i0}  (\vec p)
 -T_{0i}  (-\vec p) \right)  e^{\mathrm{i} \vec p\cdot (\vec x-\vec y)}   \,.
\end{align}
Using
\begin{eqnarray}
\left[T_{i0}  (\vec p)
 -T_{0i}  (-\vec p)\right]_{p_0=\omega_p}=-2g^2\omega_pp_i  \,,
\end{eqnarray}
we can write
\begin{align}
\left[\dot A_i({t,\vec {x}}), \dot A_0 (t,{\vec {y}} )\right]&=- \int \frac{\mathrm{d}^2\vec p}
{(2\pi)^2}    \frac{ \omega_p }{2 } \left( -2g^2\omega_pp_i  \right) e^{\mathrm{i} \vec p\cdot (\vec x-\vec y)}
\nonumber \\
&\phantom{{}={}}+\int \frac{\mathrm{d}^2\vec p}
{(2\pi)^2}    \frac{\Omega_p }{2  } \left( -2g^2\Omega_pp_i \right)  e^{\mathrm{i} \vec p\cdot (\vec x-\vec y)}   \,.
\end{align}
 Therefore,
\begin{align}
\left[  \dot  A_{i}(t, \vec x),   \dot A_{0}(t, \vec y)  \right]&=
\int \frac{\mathrm{d}^2\vec p}
{(2\pi)^2}  g^2p_i  (\omega_p^2-\Omega_p^2) e^{\mathrm{i} \vec p\cdot (\vec x-\vec y)} \nonumber \\
&=
- \int \frac{\mathrm{d}^2\vec p}
{(2\pi)^2}  p_i e^{\mathrm{i} \vec p\cdot (\vec x-\vec y)} \,.
\end{align}
By employing $p_i=\mathrm{i}\partial_i$, we arrive at
\begin{eqnarray}\label{comm5}
\left[  \dot  A_{i}(t, \vec x),   \dot A_{0}(t, \vec y)  \right]&=&\mathrm{i}\partial_i \delta^{(2)}(\vec x-\vec y)\,.
\end{eqnarray}
%............................................................................................................
\addtocontents{toc}{\protect\setcounter{tocdepth}{1}}
\subsection{Commutator $\boldsymbol{[\dot A_i({t,\vec{x}}), P_0 (t,{\vec {y}} )]}$}
\addtocontents{toc}{\protect\setcounter{tocdepth}{2}}
%............................................................................................................
We have
\begin{align}
\left[\dot  A_i({t,\vec{x}}), P_0( {t,\vec{y}} )\right]&=\left[\dot A_i({t,\vec{x}}),-\dot A_0(t,\vec y)\right. \notag \\
&\phantom{{}={}}\hspace{1cm}\left.-\frac{g}{2}\epsilon^{mk}\partial_m \dot A_k(t,\vec y)\right]
\nonumber \\
&=-\left[ \dot A_i({t,\vec{x}}), \dot A_0(t, \vec y)  \right] \notag \\
&\phantom{{}={}}+\frac{g}{2}\epsilon^{mk}\partial_m \left[\dot A_i({t,\vec{x}}),\dot A_k(t, \vec y) \right]\,.\nonumber \\
\end{align}
These commutators have been found in subsections~\ref{SubC4}, \ref{SubC5}
and after inserting their results we obtain
\begin{align}
\left[\dot A_i({t,\vec{x}}), P_0( {t,\vec{y}} )\right]&=\mathrm{i}\partial_i \delta^{(2)}(\vec x-\vec y) \notag \\
&\phantom{{}={}}+\frac{g}{2}\epsilon^{mk}\partial_m \left(\frac{-\mathrm{i}}{g} \epsilon_{ik} \delta^{(2)}(\vec x-\vec y)\right )\,.
\end{align}
Therefore,
\begin{equation}
\left[\dot  A_i({t,\vec{x}}), P_0( {t,\vec{y}} )\right]=\frac{\mathrm{i}}{2} \partial_i   \delta^{(2)}(\vec x-\vec y)\,.
\end{equation}

%............................................................................................................
\addtocontents{toc}{\protect\setcounter{tocdepth}{1}}
\subsection{Commutator $\boldsymbol{[\dot A_i({t,\vec {x}}), P^j (t,{\vec {y}} )]}$}
\addtocontents{toc}{\protect\setcounter{tocdepth}{2}}
%............................................................................................................
Consider
\begin{align}
&\left[\dot A_i({t,\vec {x}}), P^j (t,{\vec {y}} )\right]=\left[ \dot A_i(t,\vec x),
 -\dot A^j(t,{\vec {y}} )\nonumber \right.  \\
&\phantom{{}={}}\left. +\frac{g}{2}  \epsilon^{jk}\Box  A_k
(t,{\vec {y}} )+\frac{g}{2}  \epsilon^{jk}
\ddot A_k (t,{\vec {y}} )-\frac{g}{2}  \epsilon^{jk} \partial_k \dot A_0 (t,{\vec {y}} )\right]\,.
\end{align}
Hence,
\begin{align}
\left[\dot A_i({t,\vec {x}}), P^j (t,{\vec {y}} )\right]&=- \left[ \dot A_i(t,\vec x),
 \dot A^j(t,{\vec {y}} ) \right]
 \nonumber  \\
&\phantom{{}={}}+\left[ \dot A_i(t,\vec x), \frac{g}{2}  \epsilon^{jk}\Box  A_k
(t,{\vec {y}} )  \right]   \nonumber  \\
&\phantom{{}={}}+\left[ \dot A_i(t,\vec x), \frac{g}{2}
 \epsilon^{jk} \ddot A_k (t,{\vec {y}} )\right]  \nonumber  \\
&\phantom{{}={}} -\left[ \dot A_i(t,\vec x), \frac{g}{2}
  \epsilon^{jk} \partial_k \dot A_0 (t,{\vec {y}} )\right]\,.
\end{align}
We get
\begin{align}
\left[\dot A_i({t,\vec {x}}), P^j (t,{\vec {y}} )\right]&=\frac{\mathrm{i}}{g} \epsilon_{i}^{\phantom{i}j}
\delta^{(2)}(\vec x-\vec y) \nonumber  \\
&\phantom{{}={}}- \frac{1}{2g}  \epsilon^{jk}  \left[ \dot G_i(t,\vec x),   G_k
(t,{\vec {y}} )  \right]   \nonumber  \\
&\phantom{{}={}}+\frac{g}{2}  \epsilon^{jk} \left[ \dot A_i(t,\vec x),
\ddot A_k (t,{\vec {y}} )\right]   \nonumber  \\
&\phantom{{}={}}+ \frac{g}{2}  \epsilon^{jk} \partial_k \left[ \dot A_i(t,\vec x),
  \dot A_0 (t,{\vec {y}} )\right] \,,
\end{align}
where we have also used~Eq.~(\ref{comm4}).

Since
\begin{align}
& \left[ \dot G_i(t,\vec x),   G_k
(t,{\vec {y}} )  \right]=-\mathrm{i}(\eta_{ik}   +g^2\partial_i \partial_k)   \delta^{(2)}(\vec x-\vec y)\,,\nonumber\\
& \left[ \dot A_i(t,\vec x), \ddot A_k (t,{\vec {y}} )\right] =\frac{\mathrm{i}}{g^2}(\eta_{ik}
  +g^2\partial_i \partial_k)   \delta^{(2)}(\vec x-\vec y)\,,
\end{align}
and with~Eq.~(\ref{comm5}), we write
\begin{align}
&\left[\dot A_i({t,\vec {x}}), P^j (t,{\vec {y}} )\right]\nonumber \\ &=\frac{\mathrm{i}}{g} \epsilon_{i}^{\phantom{i}j} \delta^{(2)}(\vec x-\vec y)
 \notag \\
&\phantom{{}={}}- \frac{1}{2g}  \epsilon^{jk}  \left[-\mathrm{i}  \left(\eta_{ik}   +g^2\partial_i \partial_k \right)   \delta^{(2)}(\vec x-\vec y)\right]
  \notag\\
&\phantom{{}={}}+ \frac{g}{2}  \epsilon^{jk} \left[ \frac{\mathrm{i}}{g^2}(\eta_{ik}   +g^2\partial_i \partial_k)
  \delta^{(2)}(\vec x-\vec y)\right] \notag \\
&\phantom{{}={}}+ \frac{g}{2}  \epsilon^{jk} \partial_k \left[
    -\mathrm{i}\partial_i \delta^{(2)}(\vec x-\vec y)  \right] \,.
\end{align}
We see that the first, second, and fourth term cancel and are left with the result
\begin{align}
&\left[\dot A_i({t,\vec {x}}), P^j (t,{\vec {y}} )\right]=
\frac{\mathrm{i}g}{2}  \epsilon^{jk} \partial_i  \partial_k \delta^{(2)}(\vec x-\vec y) \,.
\end{align}
%............................................................................................................
\addtocontents{toc}{\protect\setcounter{tocdepth}{1}}
\subsection{Commutator $\boldsymbol{[\dot A_i({t,\vec {x}}), \Pi^j (t,{\vec {y}} )]}$}
\addtocontents{toc}{\protect\setcounter{tocdepth}{2}}
%............................................................................................................
Inserting the field operators, we have
\begin{align}
\left[\dot A_i({t,\vec {x}}), \Pi^j (t,{\vec {y}} )\right]&= \left[ \dot A_i(t,\vec x),  -\frac{g}{2}
  \epsilon^{jk} \dot A_k (t,{\vec {y}} )    \right.   \nonumber \\
&\phantom{{}={}}\left.\,+\frac{g}{2}  \epsilon^{jk} \partial_k  A_0 (t,{\vec {y}} )  \right]\,,
\end{align}
which is equal to
\begin{align}
\left[\dot A_i({t,\vec {x}}), \Pi^j (t,{\vec {y}} )\right]&=-\frac{g}{2}  \epsilon^{jk} \left[
 \dot A_i(t,\vec x),   \dot A_k (t,{\vec {y}} )\right]   \nonumber \\
&\phantom{{}={}} -\frac{g}{2}  \epsilon^{jk} \partial_k
  \left[   \dot A_i(t,\vec x),  A_0 (t,{\vec {y}} )  \right]\,.
\end{align}
The second commutator is zero and after using~Eq.~(\ref{comm4}) we find
\begin{align}
&\left[\dot A_i({t,\vec {x}}), \Pi^j (t,{\vec {y}} )\right]=-\frac{g}{2}  \epsilon^{jk} \left
[ -\frac{\mathrm{i}}{g} \epsilon_{ik} \delta^{(2)}(\vec x-\vec y) \right]\,.
\end{align}
Therefore, our result is
\begin{eqnarray}
\left[\dot A_i({t,\vec {x}}), \Pi^j (t,{\vec {y}} )\right]&=&\frac{\mathrm{i}}{2}  \delta^{ij}
 \delta^{(2)}(\vec x-\vec y)\,.
\end{eqnarray}
%............................................................................................................
\addtocontents{toc}{\protect\setcounter{tocdepth}{1}}
\subsection{Commutator $\boldsymbol{[\dot A_0({t,\vec {x}}), P^i (t,{\vec {y}} )]}$}
\addtocontents{toc}{\protect\setcounter{tocdepth}{2}}
%............................................................................................................
Here we compute one commutator which gives zero. We start with
\begin{align}
\left[\dot A_0({t,\vec {x}}), P^i(t,{\vec {y}} )\right]&=\left[ \dot A_0(t,\vec x),
-\dot A^i(t,{\vec {y}} )+\frac{g}{2}  \epsilon^{ij}\Box  A_j
(t,{\vec {y}} )\nonumber \right.  \\
&\phantom{{}={}}\left.+\,\frac{g}{2}  \epsilon^{ik} \ddot
A_k (t,{\vec {y}} )-\frac{g}{2}  \epsilon^{ik} \partial_k \dot A_0 (t,{\vec {y}} )\right]\,,
\end{align}
which yields
\begin{align}
\left[\dot A_0({t,\vec {x}}), P^i(t,{\vec {y}} )\right]&=- \left[ \dot A_0({t,\vec {x}}),
 \dot A^i (t,{\vec {y}} )\right] \notag \\
&\phantom{{}={}}-\frac{1}{2g}  \epsilon^{ik}   \left[ \dot G_0(t,\vec x),   G_k
(t,{\vec {y}} ) \right]   \nonumber\\
&\phantom{{}={}}+ \frac{g}{2}  \epsilon^{ik} \left[ \dot
 A_0({t,\vec {x}}) ,\ddot A_k (t,{\vec {y}} )  \right] \nonumber\\
&\phantom{{}={}}+\frac{g}{2}  \epsilon^{ik} \partial_k   \left[ \dot A_0({t,\vec {x}}),  \dot A_0 (t,{\vec {y}} )\right]\,.
\end{align}
The last term is zero and so
\begin{align}
\left[\dot A_0({t,\vec {x}}), P^i(t,{\vec {y}} )\right]&=- \left[ \dot A_0({t,\vec {x}}),
 \dot A^i (t,{\vec {y}} )\right] \nonumber\\
&\phantom{{}={}} -\frac{1}{2g}  \epsilon^{ik}   \left[ \dot G_0(t,\vec x),   G_k
(t,{\vec {y}} ) \right]   \nonumber\\
&\phantom{{}={}}+ \frac{g}{2}  \epsilon^{ik} \left[ \dot
 A_0({t,\vec {x}}) ,\ddot A_k (t,{\vec {y}} )  \right]  \,.
\end{align}
 We need the three elements
\begin{subequations}
\begin{align}
&\left[ \dot A_0({t,\vec {x}}),
 \dot A_i (t,{\vec {y}} )\right]  =- \mathrm{i}\partial_i \delta^{(2)}(\vec x-\vec y)  \,,
\\ &
\left[  \dot G_0   (t,{\vec {x}} ),
G_k (t,{\vec {y}} ) \right]  =  - \mathrm{i}  g  \epsilon_{k m}\partial^m  \, \delta^{(2)}(\vec x-\vec y) \,,
 \\ &
  \left[    \dot A_0 (t,{\vec {x}}), \ddot A_k (t,{\vec {y}} )\right] =\frac{\mathrm{i}}{g}
  \epsilon_{km} \partial^m \delta^{(2)}(\vec x-\vec y) \,.
\end{align}
\end{subequations}
Inserting the latter results gives
\begin{align}
\left[\dot A_0({t,\vec {x}}), P^i(t,{\vec {y}} )\right]&=- \left[  \mathrm{i}\partial_i
\delta^{(2)}(\vec x-\vec y)  \right] \nonumber\\
&\phantom{{}={}} -\frac{1}{2g}  \epsilon^{ik}   \left[ - \mathrm{i}
g  \epsilon_{k m}\partial^m  \, \delta^{(2)}(\vec x-\vec y) \right] \nonumber\\
&\phantom{{}={}}+\frac{g}{2}  \epsilon^{ik} \left[ \frac{\mathrm{i}}{g}
  \epsilon_{km} \partial^m \delta^{(2)}(\vec x-\vec y)   \right]\,.
\end{align}
Therefore, our result is
\begin{eqnarray}
\left[\dot A_0({t,\vec {x}}), P^i(t,{\vec {y}} )\right]&=&0\,.
\end{eqnarray}
%............................................................................................................
\addtocontents{toc}{\protect\setcounter{tocdepth}{1}}
\subsection{Commutator $\boldsymbol{[P_0({t,\vec {x}}), \Pi^i (t,{\vec {y}} )]}$}
\addtocontents{toc}{\protect\setcounter{tocdepth}{2}}
%............................................................................................................
We start with
\begin{align}
&\left[P_0({t,\vec{x}}), \Pi^i( {t,\vec{y}} )\right] \nonumber \\& =\left[ -\dot
A_0(t, \vec x)-\frac{g}{2}\epsilon^{mr}\partial_m \dot A_r(t, \vec x)  ,
-\frac{g}{2}  \epsilon^{ik} \dot A_k (t,{\vec {y}} ) \right.  \nonumber \\
&\phantom{{}={}}\left.+\,\frac{g}{2}  \epsilon^{ik}
\partial_k  A_0 (t,{\vec {y}} ) \right] \,,
\end{align}
or, which is the same,
\begin{align}
\left[P_0({t,\vec{x}}), \Pi^i( {t,\vec{y}} )\right]&=\frac{g}{2}  \epsilon^{ik}
 \left[ \dot A_0({t,\vec{x}}),  \dot A_k (t,{\vec {y}} )  \right]
\nonumber   \\
&\phantom{{}={}}+\frac{g}{2}  \epsilon^{ik} \partial_k \left[ \dot A_0({t,\vec{x}}),
A_0 (t,{\vec {y}} ) \right]\nonumber   \\
&\phantom{{}={}}+\frac{g^2}{4}\epsilon^{mr}
 \partial_m \epsilon^{ik}  \left[  \dot A_r(t, \vec y) ,  \dot A_k (t,{\vec {y}} ) \right]
\,.
\end{align}
Hence, from the previous results of Eqs.~(\ref{comm5}), (\ref{A_0}), and (\ref{comm4}) one has
\begin{align}
&\left[P_0({t,\vec{x}}), \Pi^i( {t,\vec{y}} )\right] \nonumber \\&=\frac{g}{2}
  \epsilon^{ik} \left[-\mathrm{i}\partial_k \delta^{(2)}(\vec x-\vec y)\right]
+\frac{g}{2}  \epsilon^{ik} \partial_k \left[\mathrm{i}\delta^{(2)}(\vec x-\vec y)
\right]\nonumber   \\
&\phantom{{}={}}+\frac{g^2}{4}\epsilon^{mr} \partial_m \epsilon^{ik}
 \left[  \frac{-\mathrm{i}}{g} \epsilon_{rk} \delta^{(2)}(\vec x-\vec y)  \right]
\,.
\end{align}
The first and second term cancel each other and we arrive at
\begin{eqnarray}
\left[P_0({t,\vec{x}}), \Pi^i( {t,\vec{y}} )\right]&=& \frac{\mathrm{i}g}{4}\epsilon^{im} \partial_m
   \delta^{(2)}(\vec x-\vec y)
\,.
\end{eqnarray}
%............................................................................................................
\addtocontents{toc}{\protect\setcounter{tocdepth}{1}}
\subsection{Commutator $\boldsymbol{[\Pi_0({t,\vec {x}}), P^i (t,{\vec {y}} )]}$}
\addtocontents{toc}{\protect\setcounter{tocdepth}{2}}
%............................................................................................................
We have
\begin{align}
&\left[\Pi_0({t,\vec {x}}), P^i(t,{\vec {y}} )\right]\nonumber \\&=\left[ \frac{g}{2}  \epsilon^{lm}
\partial_l A_m (t,{\vec {x}} ), -\dot A^i(t,{\vec {y}} )+\frac{g}{2}  \epsilon^{ij}\Box  A_j
(t,{\vec {y}} )\nonumber \right.  \\
&\phantom{{}={}}\left.\,+\,\frac{g}{2}  \epsilon^{ik} \ddot
 A_k (t,{\vec {y}} )-\frac{g}{2}  \epsilon^{ik} \partial_k \dot A_0 (t,{\vec {y}} )\right]\,.
\end{align}
The only nonzero contributions are
\begin{align}
\left[\Pi_0({t,\vec {x}}), P^i(t,{\vec {y}} )\right]&=-\frac{1}{4}  \epsilon^{lm}\partial_l
  \epsilon^{ij} \left[  G_m (t,{\vec {x}} ),   G_j
(t,{\vec {y}} ) \right] \nonumber  \\
&\phantom{{}={}}+ \frac{g^2}{4}  \epsilon^{lm} \partial_l
 \epsilon^{ik}     \left[ A_m (t,{\vec {x}} ),\ddot A_k (t,{\vec {y}} )\right]  \,.
\end{align}
We need
\begin{subequations}
\begin{align}
&\left[  G_m (t,{\vec {x}} ),   G_j
(t,{\vec {y}} ) \right]  =-\mathrm{i}g\epsilon_{mj}  \delta^{(2)}(\vec x-\vec y) \,,
 \\[1ex]
 & \left[ A_m (t,{\vec {x}} ),\ddot A_k (t,{\vec {y}} )\right] =\frac{\mathrm{i}}{g}\epsilon_{mk}
    \delta^{(2)}(\vec x-\vec y) \,.
\end{align}
\end{subequations}
Inserting the previous commutators results in
\begin{align}
\left[\Pi_0({t,\vec {x}}), P^i(t,{\vec {y}} )\right]&=-\frac{1}{4}  \epsilon^{lm}\partial_l
  \epsilon^{ij} \left[  -\mathrm{i}g\epsilon_{mj}  \delta^{(2)}(\vec x-\vec y) \right]  \nonumber
  \\
&\phantom{{}={}}+\frac{g^2}{4}  \epsilon^{lm} \partial_l
 \epsilon^{ik}     \left[ \frac{\mathrm{i}}{g}\epsilon_{mk}   \delta^{(2)}(\vec x-\vec y) \right] \,,
\end{align}
and so
\begin{eqnarray}
\left[\Pi_0({t,\vec {x}}), P^i(t,{\vec {y}} )\right]&=& -\frac{\mathrm{i}g}{2}
  \epsilon^{ij}   \partial_j      \delta^{(2)}(\vec x-\vec y)   \,.
\end{eqnarray}
%............................................................................................................
\addtocontents{toc}{\protect\setcounter{tocdepth}{1}}
\subsection{Commutator $\boldsymbol{[P^i({t,\vec {x}}), P^j (t,{\vec {y}} )]}$}
\addtocontents{toc}{\protect\setcounter{tocdepth}{2}}
%............................................................................................................
Now we compute a difficult commutator, which we prove to be zero in accordance with the classical
result using the constraints and the Dirac approach.
We take advantage of the previous findings. Consider
\begin{align}
\left[P^i({t,\vec {x}}), P^j(t,{\vec {y}} )\right]&=\left[ P^i({t,\vec {x}}),
-\dot A^j(t,{\vec {y}} )+g  \epsilon^{jr}\ddot  A_r
(t,{\vec {y}} )\nonumber \right.  \\
&\phantom{{}={}}\,-\frac{g}{2}  \epsilon^{jr} \vec \nabla^2
A_r (t,{\vec {y}} ) \nonumber \\
&\phantom{{}={}}\left.\,-\,\frac{g}{2}  \epsilon^{jr} \partial_r \dot A_0 (t,{\vec {y}} )\right]\,.
\end{align}
We rewrite the latter commutator as follows:
\begin{align}
\left[P^i({t,\vec {x}}), P^j(t,{\vec {y}} )\right]&=- \left[ P^i({t,\vec {x}}) ,
\dot A^j(t,{\vec {y}} )   \right] \notag \\
&\phantom{{}={}}+  g  \epsilon^{jr}   \left[  P^i({t,\vec {x}}),  \ddot  A_r
(t,{\vec {y}} )   \right]  \notag \\
&\phantom{{}={}}- \frac{g}{2}  \epsilon^{jr} \vec \nabla^2   \left[  P^i({t,\vec {x}}),
A_r (t,{\vec {y}} )  \right]  \notag \\
&\phantom{{}={}}+\frac{g}{2}  \epsilon^{jr} \partial_r       \left [  P^i({t,\vec {x}}), \dot A_0 (t,{\vec {y}} )\right]\,.
\end{align}
The individual commutators read
\begin{subequations}
\begin{align}
\left[ P^i({t,\vec {x}}),   A_r
(t,{\vec {y}} ) \right]  &=   -\mathrm{i}  \eta^i_{\phantom{i}r} \delta^{(2)}(\vec x-\vec y) \,, \displaybreak[0]\\[1ex]
\left[   P^i({t,\vec {x}}),  \dot A^j
(t,{\vec {y}} ) \right]  &=  - \frac{\mathrm{i}g}{2}  \epsilon^{ik}\partial^j \partial_k  \delta^{(2)}(\vec x-\vec y) \,, \displaybreak[0]\\[1ex]
\left[   P^i({t,\vec {x}}),  \dot A_0
(t,{\vec {y}} ) \right]  &= 0 \,.
\end{align}
\end{subequations}
 After some calculation we also find
 \begin{eqnarray}
\left[  P^i({t,\vec {x}}), \ddot A_r(t,{\vec {y}} ) \right] =  -\mathrm{i}
\left(  \frac{ 1}{2}  \partial^i  \partial_ r+ \eta^i_{\phantom{i}r} \vec \nabla^2  \right) \delta^{(2)}(\vec x-\vec y)\,, \nonumber \\
\end{eqnarray}
Inserting all the previous contributions leads to
\begin{align}
\left[P^i({t,\vec {x}}), P^j(t,{\vec {y}} )\right]&=\frac{\mathrm{i}g}{2}
\epsilon^{ik} \partial^j  \partial_k  \delta^{(2)}(\vec x-\vec y) \nonumber \\
&\phantom{{}={}}-  \frac{\mathrm{i}g}{2}
\epsilon^{jk} \partial^i  \partial_k  \delta^{(2)}(\vec x-\vec y)\nonumber \\
&\phantom{{}={}}+\frac{\mathrm{i}g}{2}
\epsilon^{ij}\vec \nabla^2   \delta^{(2)} (\vec x-\vec y)\,.
\end{align}
Indeed, considering each case for $i,j$ separately, one can check that
\begin{align}
\left[P^i({t,\vec {x}}), P^j(t,{\vec {y}} )\right]&=0\,.
\end{align}
%............................................................................................................
\addtocontents{toc}{\protect\setcounter{tocdepth}{1}}
\subsection{Commutator $\boldsymbol{[\Pi^i({t,\vec {x}}), \Pi^j (t,{\vec {y}} )]}$}
\addtocontents{toc}{\protect\setcounter{tocdepth}{2}}
%............................................................................................................
For this last commutator we have
\begin{align}
\left[\Pi^i({t,\vec {x}}), \Pi^j (t,{\vec{y}} )\right]&=\left[  -\frac{g}{2}
 \epsilon^{im} \dot A_m (t,{\vec {x}} )+\frac{g}{2}  \epsilon^{im}
 \partial_m  A_0 (t,{\vec {x}} ) ,  \right.   \nonumber \\
&\phantom{{}={}}\left.\,-\frac{g}{2}  \epsilon^{jk} \dot
 A_k (t,{\vec {y}} )\right.   \nonumber \\
&\phantom{{}={}}\left.\,+\,\frac{g}{2}
   \epsilon^{jk} \partial_k  A_0 (t,{\vec {y}} )  \right]  \,.
\end{align}
Due to~Eq.~(\ref{comm3}), the only contribution different from zero is
\begin{align}
\left[\Pi^i({t,\vec {x}}), \Pi^j (t,{\vec {y}} )\right]=    \frac{g^2}{4}
 \epsilon^{im}  \epsilon^{jk}  \left[  \dot A_m (t,{\vec {x}} ) ,
 \dot A_k (t,{\vec {y}} )  \right]     \,,
\end{align}
and we have
\begin{align}
 \left[\Pi^i({t,\vec {x}}), \Pi^j (t,{\vec {y}} )\right]=    \frac{g^2}{4}
 \epsilon^{im}  \epsilon^{jk}  \left[ -\frac{\mathrm{i}}{g} \epsilon_{m k}
  \delta^{(2)}(\vec x-\vec y) \right]    \,.
\end{align}
Finally, we find
\begin{eqnarray}
\left[\Pi^i({t,\vec {x}}), \Pi^j (t,{\vec {y}} )\right]&=&
  - \frac{\mathrm{i}g}{4}  \epsilon^{ij}  \delta^{(2)}(\vec x-\vec y).
\end{eqnarray}
With this final result at hand, we conclude the computation of the equal-time
commutators.
%...................................................................................
\section{Dirac theory in (2+1) dimensions}\label{App:D}
%...................................................................................
\setcounter{equation}{0}
\renewcommand{\theequation}{D.\arabic{equation}}

In the current section we would like to review the properties of a
Dirac theory in $(2+1)$ dimensions that are important for our work. The latter is based on the
Lorentz algebra $\mathfrak{so}(1,2)$, which involves three generators: two boosts $K^1$,
$K^2$ and a single rotation $L^3$. We obtain the corresponding generators as
\begin{align}
&K^1=\mathrm{i}\begin{pmatrix}
0 & 1 & 0 \\
1 & 0 & 0 \\
0 & 0 & 0 \\
\end{pmatrix}\,,\quad K^2=\mathrm{i}\begin{pmatrix}
0 & 0 & 1 \\
0 & 0 & 0 \\
1 & 0 & 0 \\
\end{pmatrix}\,,   \nonumber  \quad  \\ &L^3=\mathrm{i}\begin{pmatrix}
0 & 0 & 0 \\
0 & 0 & -1 \\
0 & 1 & 0 \\
\end{pmatrix}\,.
\end{align}
The latter satisfy the algebra
\begin{equation}
[L^3,K^1]=\mathrm{i}K^2\,,\quad [L^3,K^2]=-\mathrm{i}K^1\,,\quad [K^1,K^2]=-\mathrm{i}L^3\,.
\end{equation}
By forming appropriate linear combinations of these generators,
\begin{equation}
X=K^1+\mathrm{i}K^2\,,\quad Y=-(K^1-\mathrm{i}K^2)\,,\quad Z=2L^3\,,
\end{equation}
we obtain the Lie algebra $\mathfrak{sl}(2,\mathbb{R})$:
\begin{equation}
[Z,X]=2X\,,\quad [Z,Y]=-2Y\,,\quad [X,Y]=Z\,.
\end{equation}
Therefore, we conclude that $\mathfrak{so}(1,2)\simeq\mathfrak{sl}(2,\mathbb{R})$.

The first possibility of constructing a Dirac theory in $(2+1)$ dimensions
is to work with an irreducible spinor representation for which the Dirac matrices correspond
to the Pauli matrices (multiplied by appropriate factors) and the spinors have two components
only. An alternative is to propose a reducible spinor representation with three $(4\times 4)$
Dirac matrices and four-component spinors. We follow the latter possibility and choose the
Dirac matrices as
\begin{align}
\label{eq:dirac-matrices}
& \gamma^0=\begin{pmatrix}
\sigma^3 & 0 \\
0 & -\sigma^3 \\
\end{pmatrix}\,,\quad \gamma^1=\begin{pmatrix}
-\mathrm{i}\sigma^1 & 0 \\
0 & \mathrm{i}\sigma^1 \\
\end{pmatrix}\,,\quad \nonumber  \\  & \gamma^2=\begin{pmatrix}
-\mathrm{i}\sigma^2 & 0 \\
0 & \mathrm{i}\sigma^2 \\
\end{pmatrix}\,.
\end{align}
Note that we can define generators
\begin{equation}
\tilde{Z}=\gamma^0\,,\quad \tilde{X}=\frac{1}{2}(\gamma^1+\mathrm{i}\gamma^2)\,,\quad
\tilde{Y}=-\frac{1}{2}(\gamma^1-\mathrm{i}\gamma^2)\,,
\end{equation}
that satisfy
\begin{equation}
[\tilde{Z},\tilde{X}]=2\tilde{X}\,,\quad [\tilde{Z},\tilde{Y}]=-2\tilde{Y}\,,\quad
[\tilde{X},\tilde{Y}]=\tilde{Z}\,,
\end{equation}
showing that these new generators also form a representation of
$\mathfrak{sl}(2,\mathbb{R})$. Furthermore, the Dirac matrices of
Eq.~(\ref{eq:dirac-matrices}) obey the Clifford algebra in $(2+1)$
dimensions
\begin{equation}
\{\gamma^{\mu},\gamma^{\nu}\}=2\eta^{\mu\nu}\,,
\end{equation}
where $\eta^{\mu\nu}$ is the $(2+1)$-dimensional Minkowski metric.
It is clear that all Lorentz indices run from $0\dots 2$. The Dirac
equation is now given by
\begin{equation}
\mathcal{D}(\partial)\psi=0\,,\quad \mathcal{D}(\partial)=\mathrm{i}
\partial_{\mu}\gamma^{\mu}-m=\mathrm{i}\cancel{\partial}-m\,,
\end{equation}
with the Dirac operator $\mathcal{D}(\partial)$ acting on a four-component
spinor $\psi=\psi(x)$. Transforming the Dirac equation to momentum space
provides
\begin{equation}
\mathcal{D}(p)\tilde{\psi}=0\,,\quad \mathcal{D}(p)=\cancel{p}-m\,,
\end{equation}
with the Fourier-transformed spinor $\tilde{\psi}=\psi(p)$.
The inverse $S(p)$ of the Dirac operator in momentum space (multiplied with
$\mathrm{i}$) corresponds to the propagator:
\begin{equation}
\mathrm{i}S(p)=\frac{\mathrm{i}(\cancel{p}+m)}{p^2-m^2}\,,\quad S(p)\mathcal{D}(p)=\mathcal{D}(p)S(p)=1\,.
\end{equation}
The Feynman propagator for fermions is obtained as usual by means of the
$\mathrm{i}\epsilon$ prescription:
\begin{equation}
\label{eq:fermion-propagator}
\mathrm{i}S^F(p)=\frac{\mathrm{i}(\cancel{p}+m)}{p^2-m^2+\mathrm{i}\epsilon}\,.
\end{equation}
Requiring that the determinant of the Dirac operator vanish for nontrivial
solutions leads to the positive fermion energy
\begin{equation}
\label{eq:fermion-dispersion-relation}
E(\vec{p})=E_p=\sqrt{\vec{p}^{\,2}+m^2}\,.
\end{equation}
Solving the Dirac equation subsequently provides the following particle spinors
\begin{align}
\label{eq:particle-spinors}
u^{(1)}&=(p^1+\mathrm{i}p^2)\begin{pmatrix}
1/\sqrt{E_p-m} \\
\mathrm{i}\sqrt{E_p-m}/(p^1-\mathrm{i}p^2) \\
0 \\
0 \\
\end{pmatrix}\,,\quad \nonumber \displaybreak[0]\\
u^{(2)}&=\begin{pmatrix}
0 \\
0 \\
(p^1-\mathrm{i}p^2)/\sqrt{E_p+m} \\
\mathrm{i}\sqrt{E_p+m} \\
\end{pmatrix}\,.
\end{align}
On the other hand, the antiparticle spinors are given by:
\begin{align}
\label{eq:antiparticle-spinors}
v^{(1)}&=(p^1+\mathrm{i}p^2)\begin{pmatrix}
1/\sqrt{E_p+m} \\
\mathrm{i}\sqrt{E_p+m}/(p^1-\mathrm{i}p^2) \\
0 \\
0 \\
\end{pmatrix}\,,\quad \nonumber \displaybreak[0]\\
v^{(2)}&=\begin{pmatrix}
0 \\
0 \\
(p^1-\mathrm{i}p^2)/\sqrt{E_p-m} \\
\mathrm{i}\sqrt{E_p-m} \\
\end{pmatrix}\,.
\end{align}
These spinors are normalized such that
\begin{equation}
u^{(s)\dagger}u^{(t)}=2E_p\delta^{s,t}\,,\quad v^{(s)\dagger}v^{(t)}=2E_p\delta^{s,t}\,.
\end{equation}
We define the Dirac conjugated spinors as $\overline{u}^{(s)}=u^{(s)
\dagger}\gamma^0$ and $\overline{v}^{(s)}=v^{(s)\dagger}\gamma^0$
and derive the completeness relations
\begin{subequations}
\begin{align}
\label{eq:completeness-particle-spinors}
\sum_{s} u^{(s)}(p)\overline{u}^{(s)}(p)&=\cancel{p}+m\,, \\[2ex]
\sum_{s} v^{(s)}(p)\overline{v}^{(s)}(p)&=\cancel{p}-m\,.
\end{align}
\end{subequations}
They formally correspond to those in $(3+1)$-dimensional Dirac theory.

\end{appendix}

\providecommand{\href}[2]{#2}
%\bibliography{funcred.bib}

%.............................................................................

\end{document}